\documentclass{svproc}
\usepackage{graphicx}%
\usepackage{multirow}%
\usepackage{amsmath,amssymb,amsfonts}%
\usepackage{mathrsfs}%
\usepackage[title]{appendix}%
\usepackage{xcolor}%
\usepackage{textcomp}%
\usepackage{manyfoot}%
\usepackage{booktabs}%
\usepackage{algorithm}%
\usepackage{algorithmicx}%
\usepackage{algpseudocode}%
\usepackage{listings}%
\usepackage{tabularx}
\usepackage{comment}
\usepackage{subcaption}
\usepackage{wrapfig}
\usepackage{url}

\begin{document}
\mainmatter              
\title{Group Fairness Metrics for Community Detection Methods in Social Networks}
\titlerunning{Group Fairness Metrics}  
%
\author{Elze de Vink \and Akrati Saxena} 

\institute{Leiden Institute of Advanced Computer Science,\\ Leiden University, The Netherlands\\
\email{a.saxena@liacs.leidenuniv.nl} 
}

\maketitle             

\begin{abstract}
Understanding community structure has played an essential role in explaining network evolution, as nodes join communities which connect further to form large-scale complex networks. In real-world networks, nodes are often organized into communities based on ethnicity, gender, race, or wealth, leading to structural biases and inequalities. Community detection (CD) methods use network structure and nodes' attributes to identify communities, and can produce biased outcomes if they fail to account for structural inequalities, especially affecting minority groups. In this work, we propose group fairness metrics ($\Phi^{F*}_{p}$) to evaluate CD methods from a fairness perspective. We also conduct a comparative analysis of existing CD methods, focusing on the performance-fairness trade-off, to determine whether certain methods favor specific types of communities based on their size, density, or conductance. Our findings reveal that the trade-off varies significantly across methods, with no specific type of method consistently outperforming others. The proposed metrics and insights will help develop and evaluate fair and high performing CD methods.
\keywords{Community Detection, Algorithmic Fairness, Group Fairness Metrics}
\end{abstract}

\section{Introduction}\label{intro}

In social networks, nodes are organized into communities. As Barab{\'a}si defines, ``In network science, we call a community a group of nodes that have a higher likelihood of connecting to each other than to nodes from other communities'' \cite{barabasi2014network}. Social networks have structural inequalities as the phenomenon of people joining communities is strongly driven by their ethnicity, gender, race, or wealth. These networks contain communities that vary in size, density, and their connectivity within the network based on human behavior. If such inequalities are not considered while designing Social Network Analysis (SNA) algorithms, they might lead to a biased outcome, especially for minorities \cite{saxena2024fairsna}. Community detection (CD) methods use network structure and nodes' attributes to identify communities, and if they overlook structural inequalities, they cannot efficiently identify small-size groups \cite{ghasemian2019evaluating}. Misclassifying communities can further affect the fairness of other SNA methods. For example, fairness-aware methods proposed for influence maximization \cite{Farnad2020}, influence minimization \cite{Saxena2023}, link prediction \cite{Saxena2021,saxena2022nodesim}, and centrality ranking \cite{Tsioutsiouliklis2020} use community membership to get the final fair outcome. It is crucial to consider structural inequalities while designing SNA methods to ensure fairness and mitigate bias for or against all users and groups, irrespective of their size or type. 

Ghasemian et al. \cite{ghasemian2019evaluating} compared 16 CD methods and found that the number of identified communities varies a lot across different methods. The authors did not examine the impact of these methods on different types of nodes or communities with varying sizes and densities. While many metrics exist to assess the quality of detected communities \cite{Chakraborty2016}, there is no metric to evaluate the fairness of a method, particularly in relation to bias against minority groups. Fairness is not yet well defined and studied for CD methods, given that it has a vast literature \cite{fortunato2016community}. 

In this work, we first propose fairness metrics to compute fairness for each community and use them further to propose group fairness metrics $(\Phi)$ for CD methods. Next, we perform a comparative analysis of existing community detection methods using the performance-fairness trade-off to study if high-performing methods are biased towards communities of different sizes, densities, or conductance. The size of a community is the total number of nodes, density is the ratio of the number of internal edges and total possible internal edges, and conductance is the fraction of the total edge volume that connects outside the community. We classify CD methods into six classes: (i) Optimization, (ii) Spectral, (iii) Propagation, (iv) Dynamics, (v) Representation Learning, and (vi) Miscellaneous. Experiments are performed on LFR and real-world networks. Results show that it is not the case that a specific class of methods always performs better. The performance-fairness trade-off is highly dependent on particular methods, and some of the best-performing methods identifying high-quality and fair communities include Infomap, RSC-V, RSC-K, Significance, and Walktrap.

To the best of our knowledge, this is the first work of its kind. Next, we discuss the proposed metrics, followed by experimental setup, results, and conclusion.

\section{The Proposed Group Fairness Metric ($\Phi$)}

To compute the proposed fairness metrics, we begin by mapping the ground-truth and identified communities. Next, we compute the community-wise scores using the four proposed metrics, and finally, we compute the fairness of a CD method using these scores. The complete methodology is explained below. 

\subsection{Community Mapping}
Let's assume that a given network $G(V,E)$ has $m$ ground-truth communities, defined as $C=\{c_1, c_2,..., c_{m}\}$. We apply a CD method on $G$ which returns a set of predicted communities $P$ of size $k$ defined as $P=\{p_1, p_2, ..., p_{k}\}$. To evaluate the bias in a community detection method, it is essential to assess the fairness of the detection process for each ground-truth community, i.e., the extent to which each community has been accurately identified. This is achieved by mapping the ground-truth communities to the corresponding predicted communities. \\
For mapping, perform the following steps until there is at least one ground-truth and one predicted community that is not-mapped. 
\begin{enumerate}
        \item Compute Jaccard-similarity for each pair of ground-truth and predicted communities, as follows:
        \begin{equation*}
            \label{eq:jaccard_sim}
            J(c_i, p_j) = \frac{|c_i \cap p_j|}{|c_i \cup p_j|}, \forall \; i \; \& \; j
        \end{equation*}
        \item A pair having the highest similarity score is chosen and these ground-truth and predicted communities are mapped. In case of a tie, a ground-truth community is randomly chosen and mapped.
    \end{enumerate}
After this process, if there is an unmapped ground-truth community, it is marked as completely misclassified and will be mapped with an empty set.

\subsection{Community-wise Fairness Metrics}

We propose the following metrics to evaluate how effectively a ground-truth community is represented by its corresponding predicted community, taking into account both the community's nodes and edges.

\begin{enumerate}
    \item \textbf{Fraction of Correctly Classified Nodes (FCCN):}
      A simple metric to assess how well a predicted community ($p_j$) represents the ground-truth community ($c_i$) is the fraction of ground-truth nodes present in the predicted community. The FCCN is computed as follows:
        \begin{equation}
            \textit{FCCN}(c_i, p_j) = \frac{|c_i \cap p_j|}{|c_i|}
        \end{equation}
        
    \item  \textbf{F1 Score}: The FCCN focuses mainly on common nodes in the ground-truth and predicted communities, but it does not account for extra nodes present in the predicted community. To address this, we propose the F1 score (inspired by the F1~\cite{Chinchor1992} score used in machine learning), i.e., calculated as:
        \begin{equation}\label{eq:f1}
            F1(c_i, p_j) = \frac{2|c_i \cap p_j|}{|c_i| + |p_j|}
        \end{equation}
        
    \item \textbf{Fraction of Correctly Classified Edges (FCCE):} Community structure is primarily driven by edges, and a thorough evaluation of a CD method's performance should include an analysis of community edges. The FCCE metric quantifies the number of edges from the ground-truth community that are present in the mapped predicted community. It is calculated as:
    \begin{equation}
            FCCE(c_i, p_j) = \frac{|E(c_i) \cap E(p_j)|}{|E(c_i)|}
    \end{equation}
    where $E(c_i)$ is the set of intra-community edges of a community $c_i$ and it is computed as: $E(c_i) = \{(u, v) \in E \mid u \in c_i \text{ and } v \in c_i \}$.
\end{enumerate}

\subsection{Group Fairness Metric ($\Phi$)}

\begin{wrapfigure}{r}{0.45\textwidth}
  \begin{center}
    \includegraphics[width=5cm]{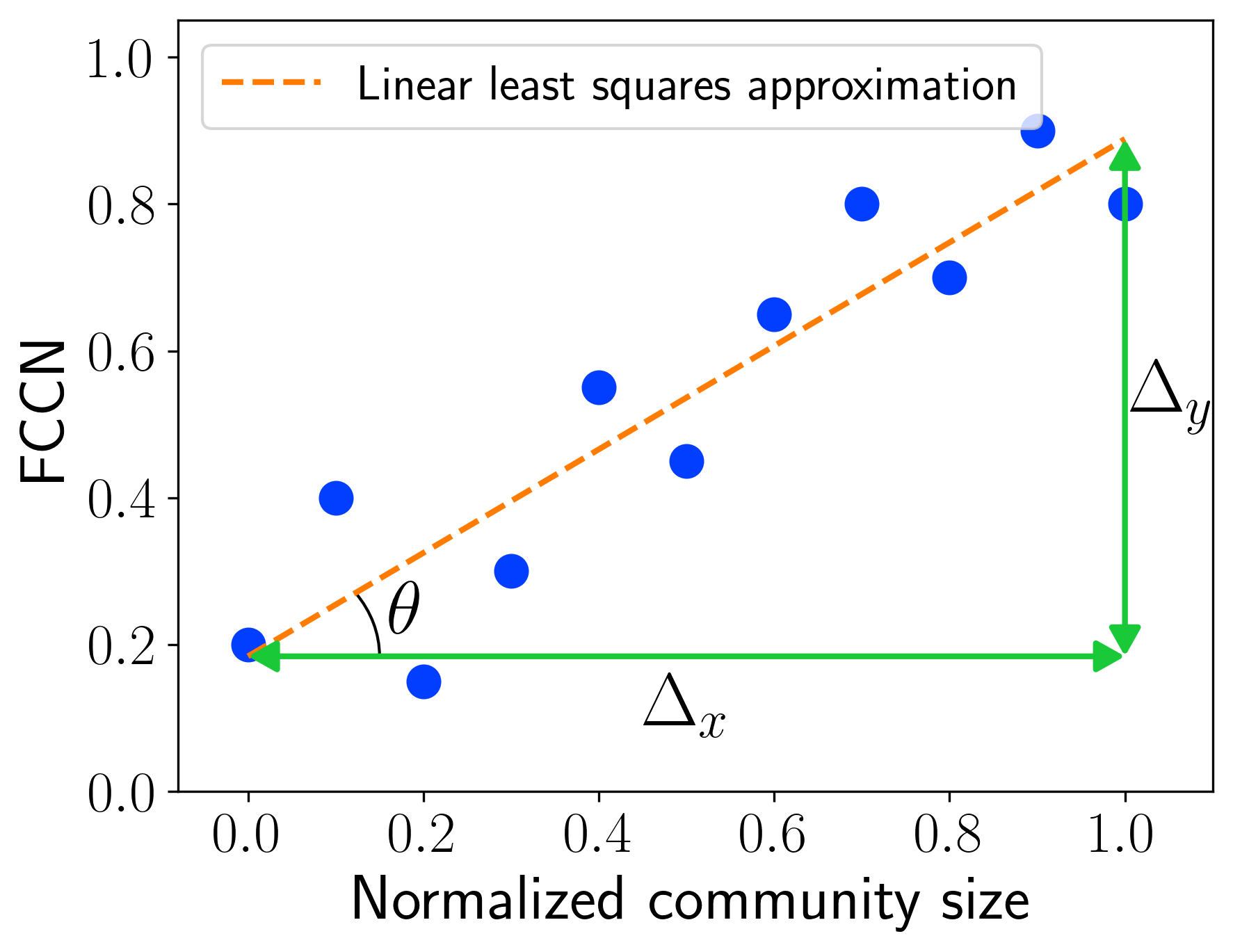}
  \end{center}
  \caption{FCCN vs. normalized community size for each community on a dummy network. The best fit line shows the trend of a CD method.}
  \label{fig:fairness-example}
\end{wrapfigure}

Our aim is to investigate whether a given CD method favors communities of specific properties based on their size, density, or conductance, using the aforementioned community-wise fairness metrics. A straightforward approach is to plot these metrics for each community against their corresponding attribute values. For example, Fig.~\ref{fig:fairness-example} shows FCCN vs. normalized community size for a dummy network, showing that larger communities are more accurately identified than smaller ones. To ensure comparability across different networks, we normalize community attribute using min-max scaling so that they range from 0 to 1.

To compute the group fairness metric ($\Phi$), we fit a linear line using least squares approximation~\cite{Hastie2009} on the proposed community-wise fairness metrics (represented by $F^*$) versus normalized community property; e.g., in Fig.~\ref{fig:fairness-example}, the dashed line is the best fit linear line on the community-wise fairness metric (FCCN) versus normalized community size. The fairness metric is measured using the slope of the line, which ranges from (-1, 1). The regression line provides $\Delta x$ and $\Delta y$, and to compute the angle $\theta$ we use the arctangent: $\textit{arctan}\left(\frac{\Delta y}{\Delta x}\right)$. This is further multiplied by $\frac{2}{\pi}$ to get the fairness metric value in $(-1,1)$ range as the arctangent angle is in radians within $(-\frac{\pi}{2}, \frac{\pi}{2})$. Finally, the fairness of a CD method, with respect to a per-community fairness metric (F*) and community property (p) is computed as:
    \begin{equation}
        \label{eq:angle}
        \Phi_p^{F*} = \frac{2}{\pi}\textit{arctan}\left(\frac{\Delta y}{\Delta x}\right) = \frac{2}{\pi}\textit{arctan}\left({\Delta y}\right)
    \end{equation}
The x-axis values are normalized using min-max scaling, so $\Delta x =1$ and can be removed. Using the example from Fig.~\ref{fig:fairness-example}, given the $\Delta y = 0.70$, we calculate its fairness score as follows:  $\Phi_{size}^{FCCN} = \frac{2}{\pi}\cdot \textit{arctan}(0.70) \approx 0.39$. \\
The proposed metric $(\Phi_p^{F*})$ ranges $(-1,1)$, where $0$ (a straight best fit line) indicates fair result, meaning all communities are identified either equally well or equally poorly. Negative and positive values indicate that the method favors communities with lower and higher property values $(p)$, respectively.

\subsection{Metric Behavior}\label{sec:metric-behavior}

\begin{figure}[t]
\centering
    \begin{subfigure}[b]{0.4\textwidth}            
            \includegraphics[width=\textwidth]{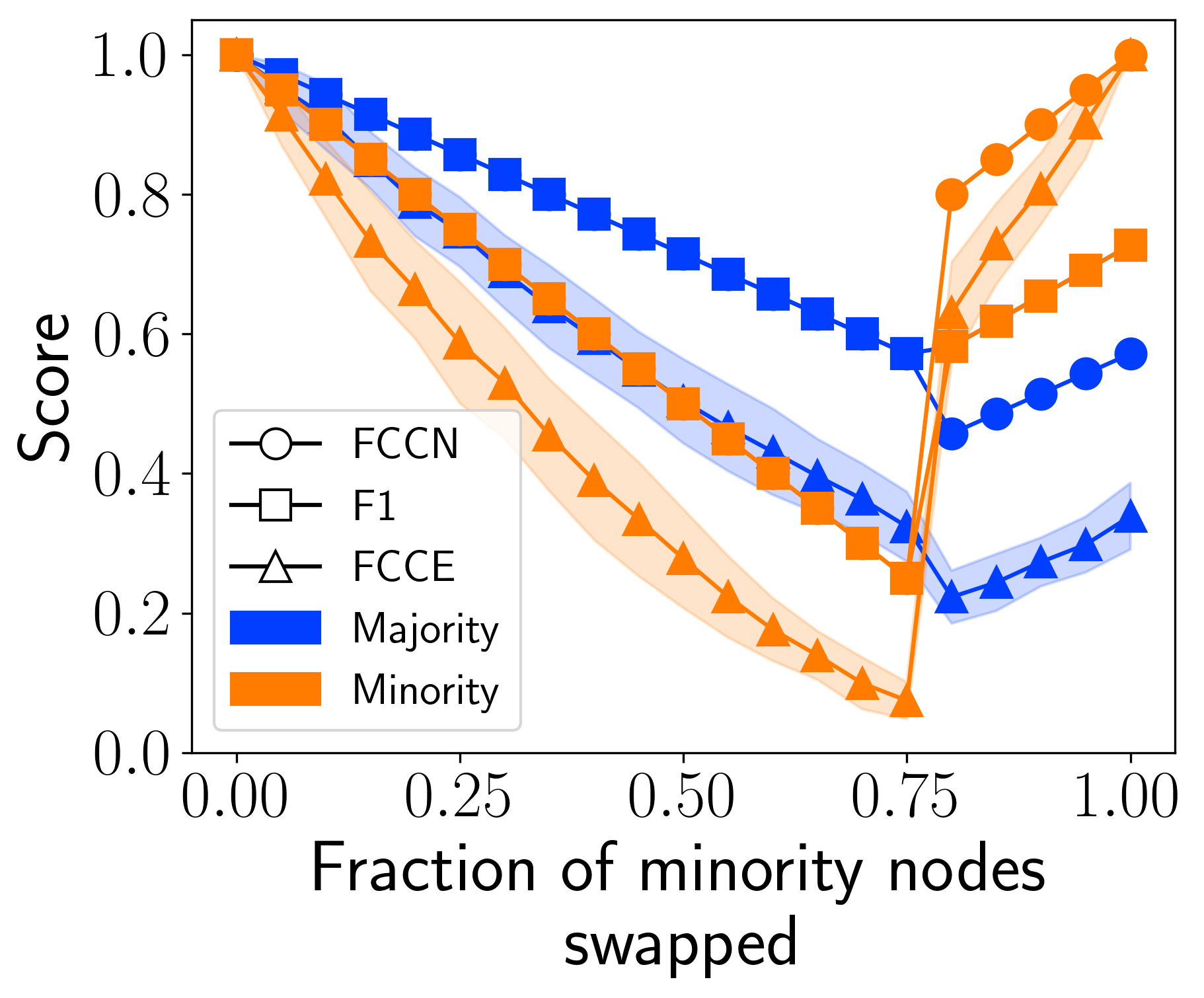}
    \end{subfigure} \quad \quad
    \begin{subfigure}[b]{0.42\textwidth}
            \centering
            \includegraphics[width=\textwidth]{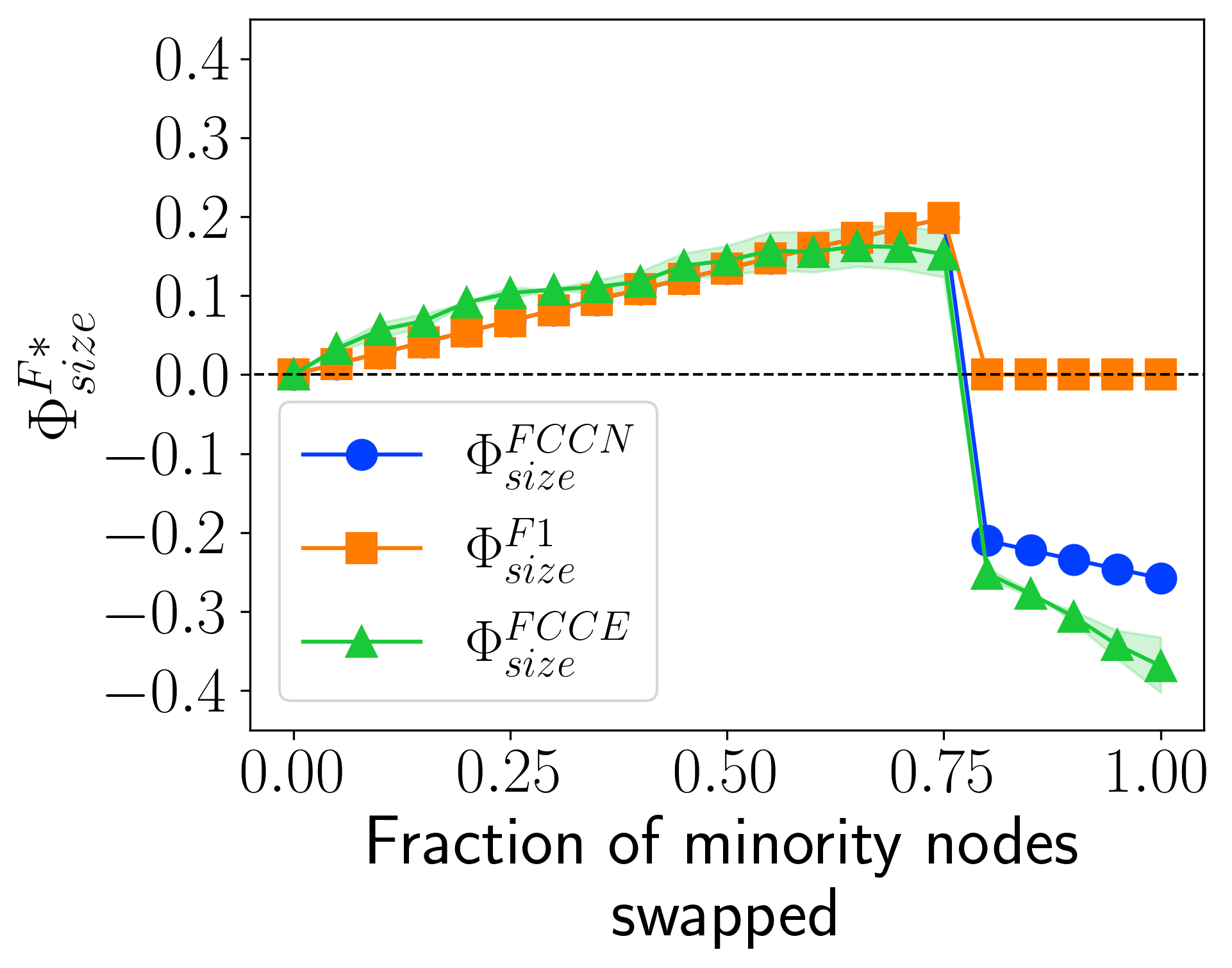}
    \end{subfigure}
    \caption{Analyzing behavior of community-wise fairness metric and group fairness on a network having minority and majority community.}
    \label{fig_metric_behavior}
\end{figure}
    
To study the behavior of the proposed metric, we create a homophilic network (homophily factor 0.9) using HICH-BA model \cite{Saxena2023} having two communities: majority (70 nodes) and minority (40 nodes), and $\sim$900 edges. Initially, predicted communities are the same as the ground truth. We then swap nodes between minority and majority from 0 to 40, and map the resulted predicted and ground-truth communities. Fig.~\ref{fig_metric_behavior} shows community-wise fairness (left) and group-fairness (right) scores with respect to swapped nodes. FCCE score can vary based on which nodes are moved, and therefore, we show the average and deviation over 20 iterations. FCCE is lower than FCCN and F1 due to homophilic characteristic of the network. 

As we switch an equal number of nodes between minority and majority, the community-wise fairness is always lower for minority than majority till a certain point ($\sim$0.75), but after that, the mapping switches between majority and minority and the fairness for minority increases as compared to the majority. Similar insights are conveyed by $\Phi^{F*}_{size}$ that initially favors the majority, but once the mapping is switched, the method either favors the minority or remains unbiased. The $\Phi^{F1}_{size}$ is fair after that point, as it also considers the nodes from the predicted communities that are not present in the ground-truth. Just to clarify, this example creates extreme conditions that might not happen in real-world. 

The proposed fairness metric ($\Phi$) offers an understanding of biases, ensuring that CD methods are evaluated not just on their overall performance but also on their fairness towards communities of different sizes, densities, and conductance.

\section{Experimental Setup}

\subsection{Community Detection Methods}
We compare 24 CD methods (refer Table~\ref{tab:cdm table}) falling under the following six classes.
\begin{enumerate}
    \item \textbf{Optimization} methods aim to optimize a quality function that describes the quality of the given partition. All methods considered in this category optimize modularity~\cite{Newman2004} except the Significance~\cite{Traag2013}. 

    \item \textbf{Dynamics} methods infer communities by analyzing traversal on the network, typically through random walks as they tend to remain within communities due to their dense connectivity compared to rest of the network. 
    
    \item \textbf{Spectral} methods identify communities using the spectral properties of matrices representing the network, such as adjacency or Laplacian matrix~\cite{fortunato2016community}. 
    
    \item \textbf{Propagation} methods utilize the iterative propagation of community labels, where each node's label is updated based on its neighbors' labels, until a stable configuration is reached representing communities in the network.

    \item \textbf{Representational} Learning methods first generate a network embedding, and then apply clustering algorithm to create partition of the network \cite{arya2022node}.

    \item \textbf{Miscellaneous} category contains the methods that do not fit to any of the previously mentioned category due to their underlying working mechanism. 
\end{enumerate}

\begin{table}[t]
    \centering
    \caption{Overview of CD methods used in experimentation from 6 classes.}
    \label{tab:cdm table}
    \begin{tabularx}{\linewidth}{@{}|X|X|X|@{}}
        \hline \multicolumn{1}{|c|}{Optimization} & \multicolumn{1}{c|}{Spectral Properties} & \multicolumn{1}{c|}{Representational} \\ \hline
        \begin{tabular}[c]{@{}l@{}}
            $\bullet$~Clauset-Newman-Moore \\\hspace{9.91664pt}Algorithm (CNM)~\cite{Clauset2004} \\
            $\bullet$~Combo~\cite{Sobolevsky2014} \\ 
            $\bullet$~Leiden~\cite{Traag2019} \\ 
            $\bullet$~Louvain~\cite{Blondel2008} \\ 
            $\bullet$~Paris~\cite{Bonald2018} \\ 
            $\bullet$~Reichardt-Bornholdt - \\\hspace{9.91664pt}configuration null model \\\hspace{9.91664pt}(RB-C)~\cite{Reichardt2006} \\ 
            $\bullet$~Reichardt-Bornholdt - \\\hspace{9.91664pt}Erdős-Rényi null model \\\hspace{9.91664pt}(RB-ER)~\cite{Reichardt2006} \\
            $\bullet$~Significance~\cite{Traag2013}
        \end{tabular}  & 
        \begin{tabular}[c]{@{}l@{}}
            $\bullet$~Eigenvector~\cite{Newman2006} \\
            $\bullet$~Regularized Spectral \\\hspace{9.91664pt}Clustering with k-means \\\hspace{9.91664pt}(RSC-K)~\cite{Zhang2018} \\
            $\bullet$~RSC sklearn Spectral \\\hspace{9.91664pt}Embedding (RCS-SSE)\\\hspace{9.91664pt}\cite{Zhang2018} \\
            $\bullet$~RSC - Vanilla (RSC-V)\\\hspace{9.91664pt}\cite{Zhang2018} \\
            $\bullet$~Spectral Clustering~\cite{Higham2007} \\ \\ \\ \\
        \end{tabular} &
        \begin{tabular}[c]{@{}l@{}}
            $\bullet$~Deepwalk~\cite{Perozzi2014} \\
            $\bullet$~Fairwalk~\cite{Rahman2019} \\
            $\bullet$~Node2Vec~\cite{Grover2016} \\ \\ \\ \\ \\ \\ \\ \\ \\ \\ \\
        \end{tabular} \\ \hline
    \end{tabularx} \\
    
    \vspace{.1cm}
    
    \begin{tabularx}{\linewidth}{@{}|X|X|X|@{}}
        \hline \multicolumn{1}{|c|}{Dynamics} & \multicolumn{1}{c|}{Propagation} & \multicolumn{1}{c|}{Miscellaneous} \\ \hline
        \begin{tabular}[c]{@{}l@{}}
            $\bullet$~Infomap~\cite{Rosvall2008} \\
            $\bullet$~Spinglass~\cite{Reichardt2006} \\ 
            $\bullet$~Walktrap~\cite{Pons2005} \\ \\ \\
        \end{tabular} &
        \begin{tabular}[c]{@{}l@{}}
            $\bullet$~Fluid~\cite{Parés2017} \\ 
            $\bullet$~Label Propagation~\cite{Cordasco2011} \\ \\ \\ \\
        \end{tabular} &
        \begin{tabular}[c]{@{}l@{}}
            $\bullet$~Expectation-Maximization \\\hspace{9.91664pt}(EM)~\cite{Newman2007} \\ 
            $\bullet$~Stochastic Block Model \\\hspace{9.91664pt}(SBM)~\cite{Peixoto2014} \\
            $\bullet$~SBM - Nested~\cite{Peixoto2014a}
        \end{tabular} \\ \hline
    \end{tabularx}
    
\end{table}

We use CDlib implementation (\url{https://cdlib.readthedocs.io/}) with default parameters for all CD methods. The code for our experiments and metric computations is available at:\\
\url{https://github.com/akratiiet/Group-Fairness-Metrics-for-Community-Detection}.

\subsection{Datasets}

\textbf{Synthetic Networks:} We use LFR benchmark model \cite{Lancichinetti2008} to generate networks of varying mixing parameter $\mu$ having 10,000 nodes, and the power law exponents of degree and community size to be 2 and 2.5, respectively.\\ 
\textbf{Real-world Networks:} These are described in Table~\ref{tab:real-world data}.

\begin{table}[]
\centering
\caption{Real-world datasets. $|C|$: total number of communities, $|c_{max}|$: size of largest community, $|c_{min}|$: size of smallest community.}
\label{tab:real-world data}
\begin{tabular}{|l|*{8}{r|}}
    \hline Dataset & \multicolumn{1}{c|}{$|V|$} & \multicolumn{1}{c|}{$|E|$} & \multicolumn{1}{c|}{\textit{Avg deg}} & \multicolumn{1}{c|}{$\textit{Max deg}$} & \multicolumn{1}{c|}{$|C|$} &\multicolumn{1}{c|}{$|c_{min}|$} & \multicolumn{1}{c|}{$|c_{max}|$}  \\ \hline 
    Polbooks~\cite{Krebs} & 105 & 441 & 8.40 & 25 & 3 & 13 & 49  \\
    Football~\cite{Girvan2002} & 115 & 613 & 10.66 & 12 & 12 & 5 & 13 \\
    Eu-core~\cite{Leskovec2014} & 986 & 16,687 & 33.85 & 347 & 42 & 1  & 107 \\ 
    \hline    
\end{tabular}
\end{table}

\section{Results}\label{results}

We study the fairness of CD methods with respect to community size, density and conductance versus the quality of identified communities, i.e., measured using the Normalized Mutual Information (NMI) \cite{Fred2003}.

\subsection{Group Fairness-Performance trade-off versus Community Size}

\begin{figure}[t]
\centering
\begin{subfigure}[b]{0.98\textwidth}            
    \includegraphics[width=\textwidth]{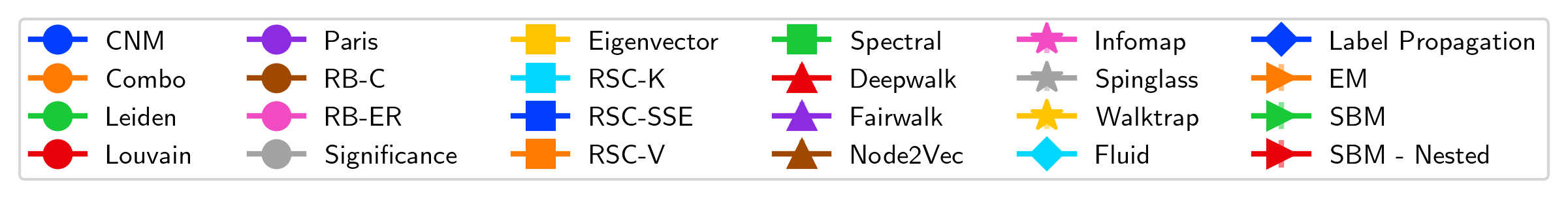}
\end{subfigure}\\
\begin{subfigure}[c]{0.05\textwidth}
\caption*{\rotatebox{90}{$\mu=0.2$}}
\end{subfigure}%
\begin{minipage}[c]{0.95\textwidth}
\includegraphics[width=0.31\textwidth]{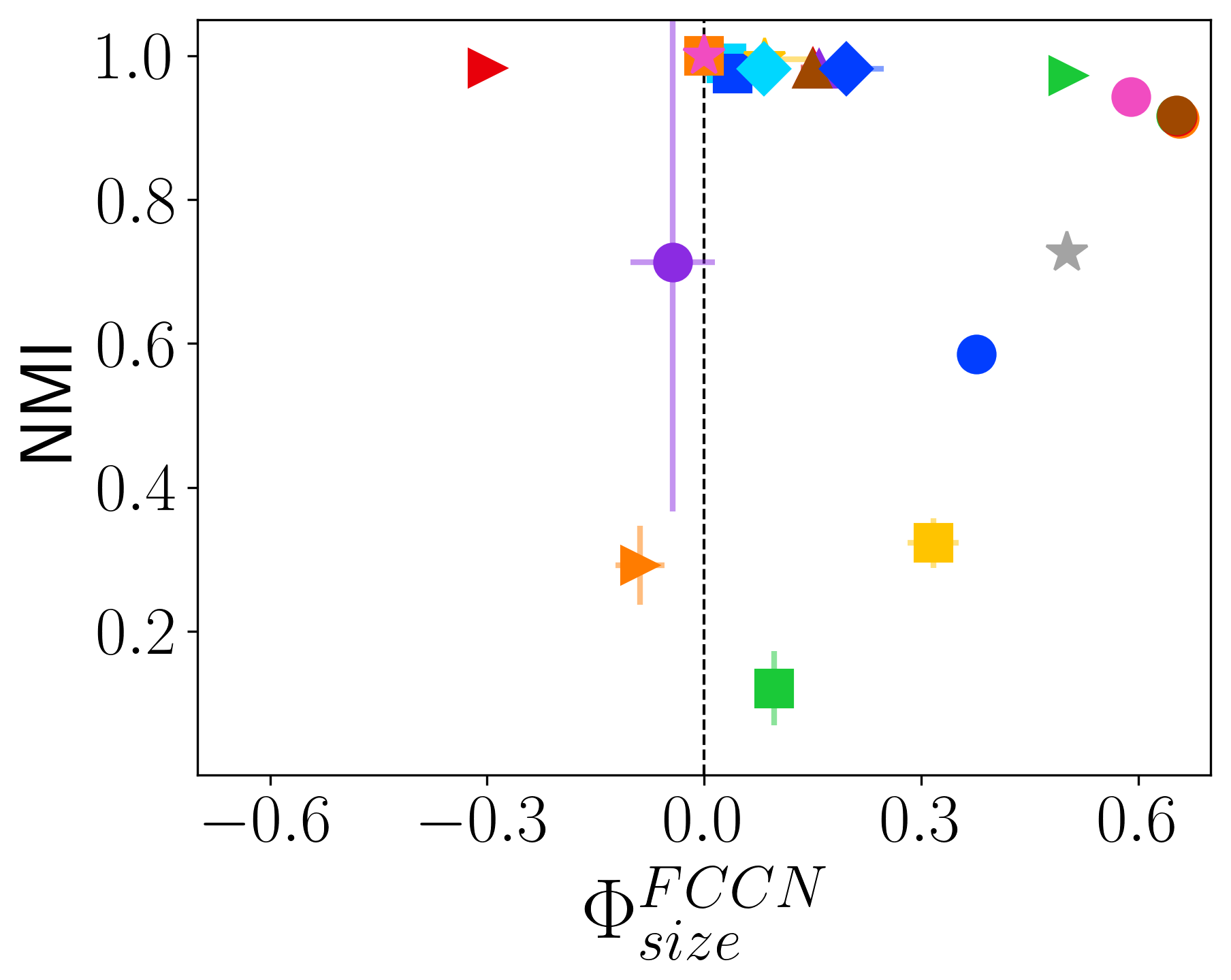}\quad
\includegraphics[width=0.31\textwidth]{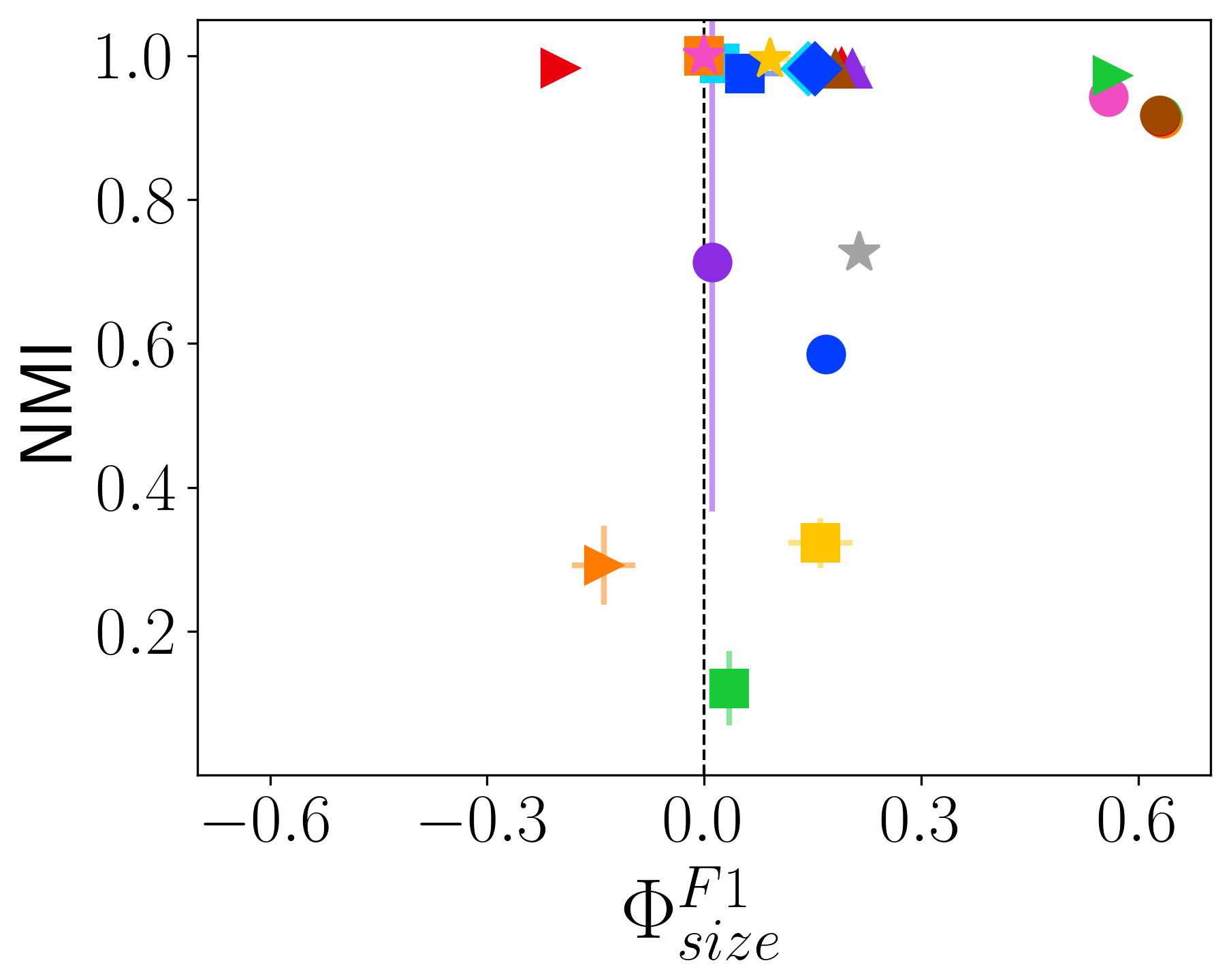}\quad
\includegraphics[width=0.31\textwidth]{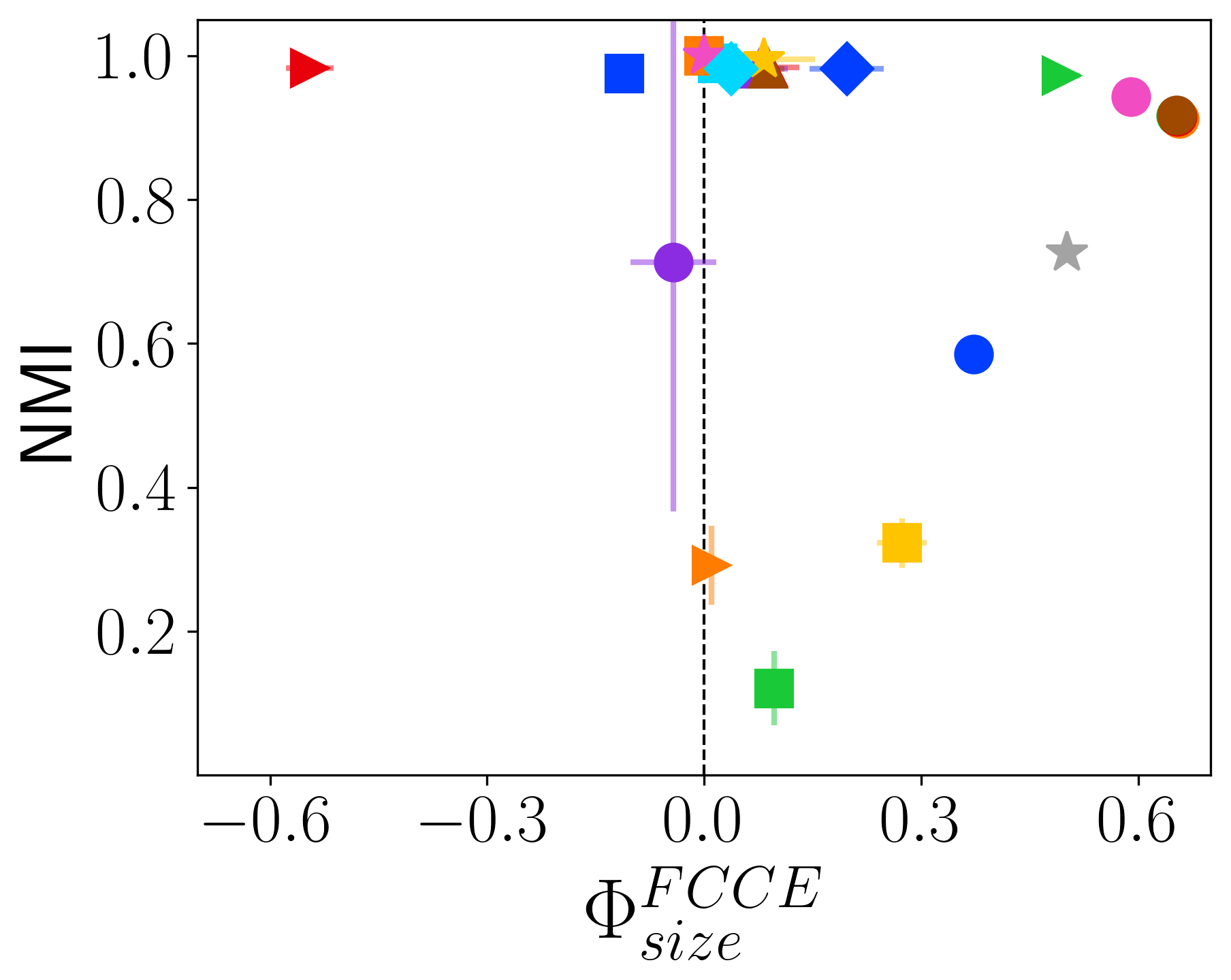}
\end{minipage}
\\
\begin{subfigure}[c]{0.05\textwidth}
\caption*{\rotatebox{90}{$\mu=0.4$}}
\end{subfigure}%
\begin{minipage}[c]{0.95\textwidth}
\includegraphics[width=0.31\textwidth]{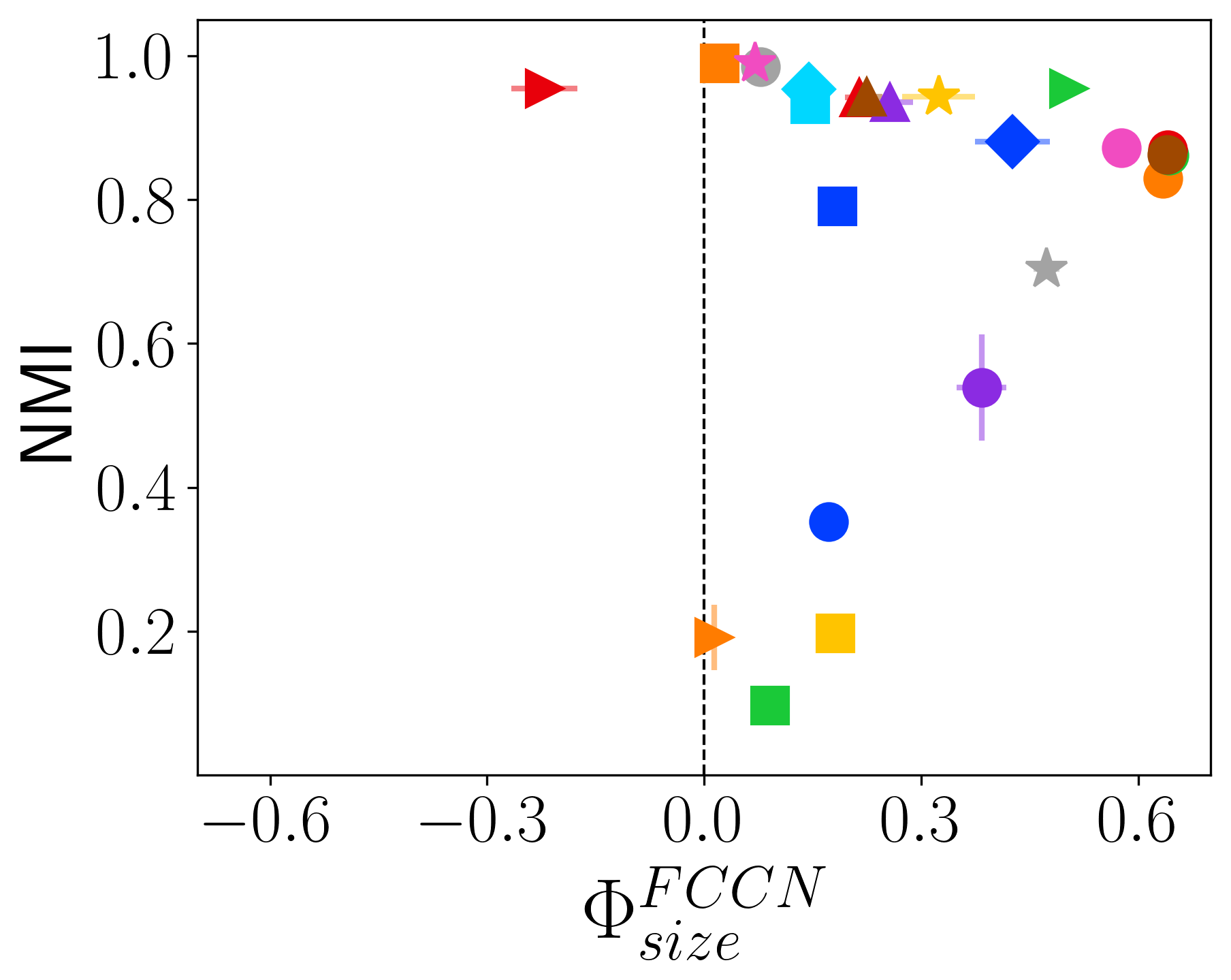}\quad
\includegraphics[width=0.31\textwidth]{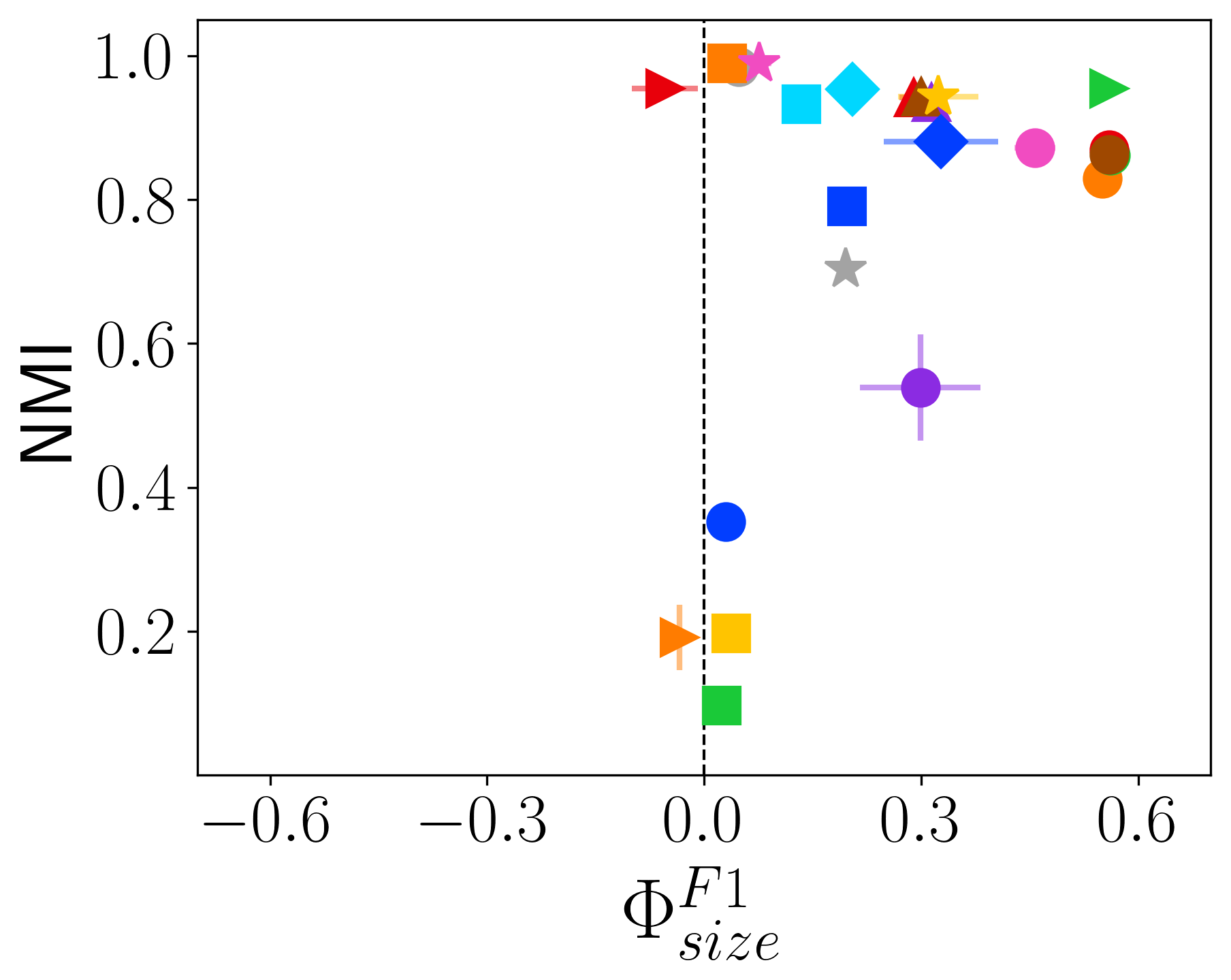}\quad
\includegraphics[width=0.31\textwidth]{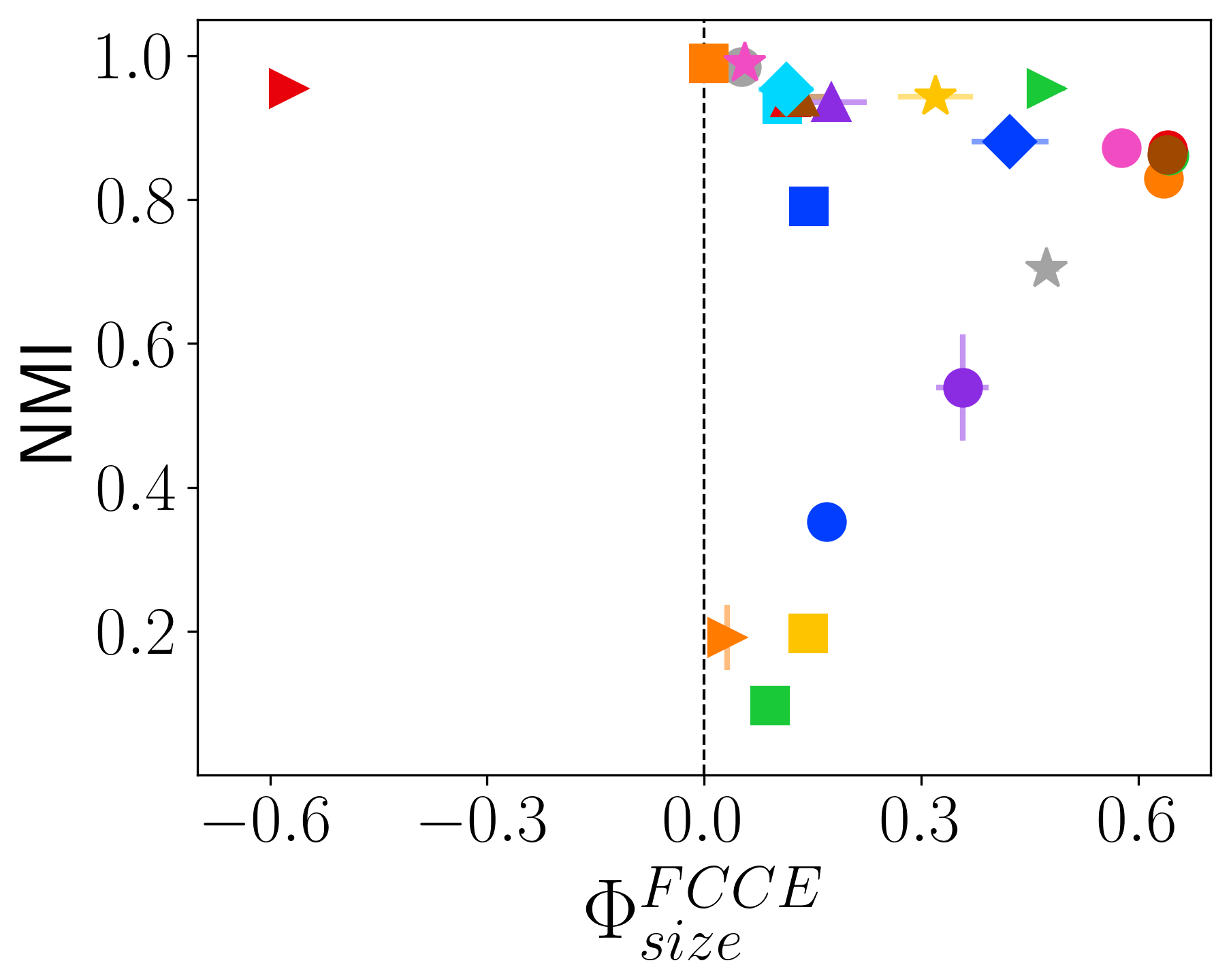}
\end{minipage}
\\
\begin{subfigure}[c]{0.05\textwidth}
\caption*{\rotatebox{90}{$\mu=0.6$}}
\end{subfigure}%
\begin{minipage}[c]{0.95\textwidth}
\includegraphics[width=0.31\textwidth]{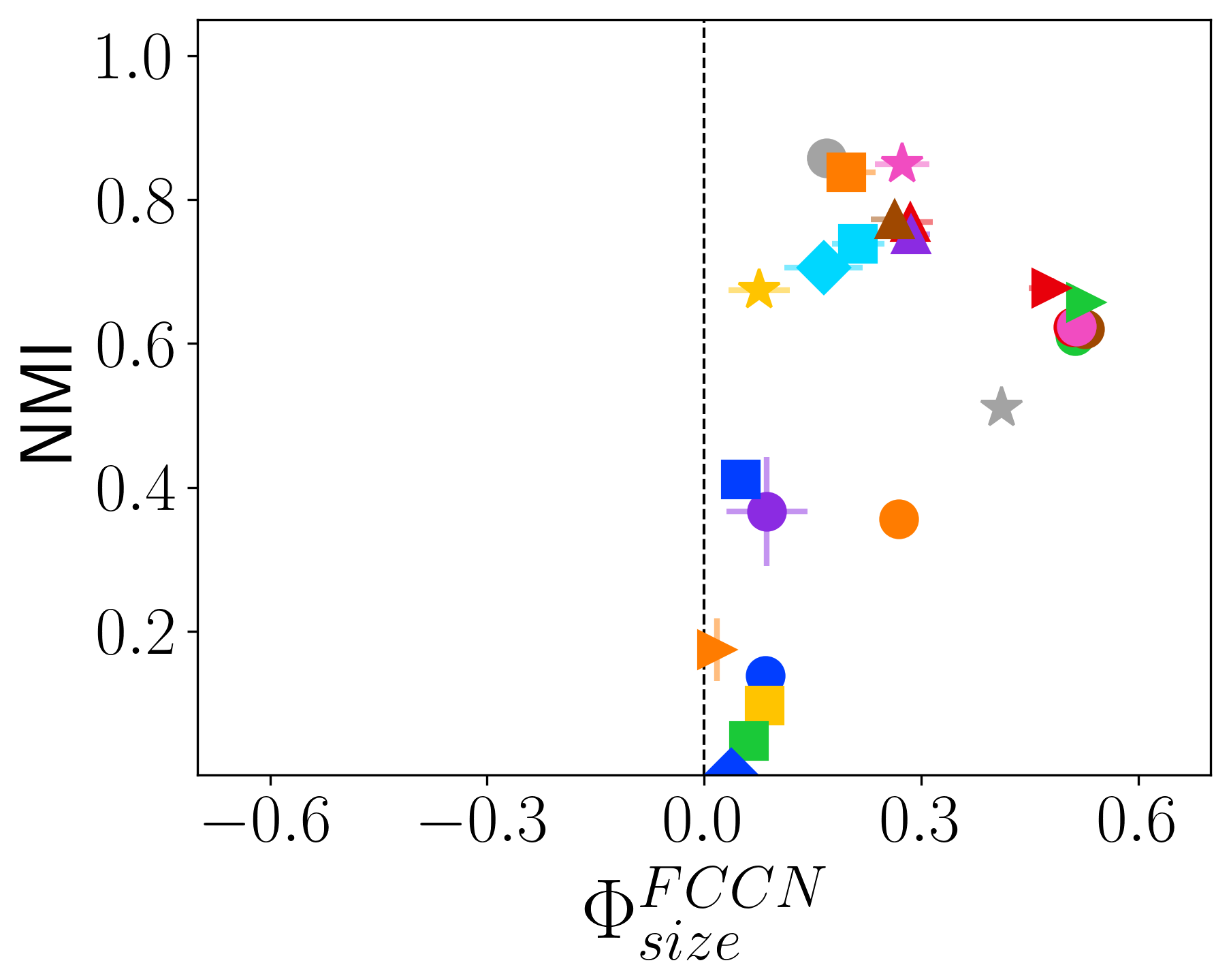}\quad
\includegraphics[width=0.31\textwidth]{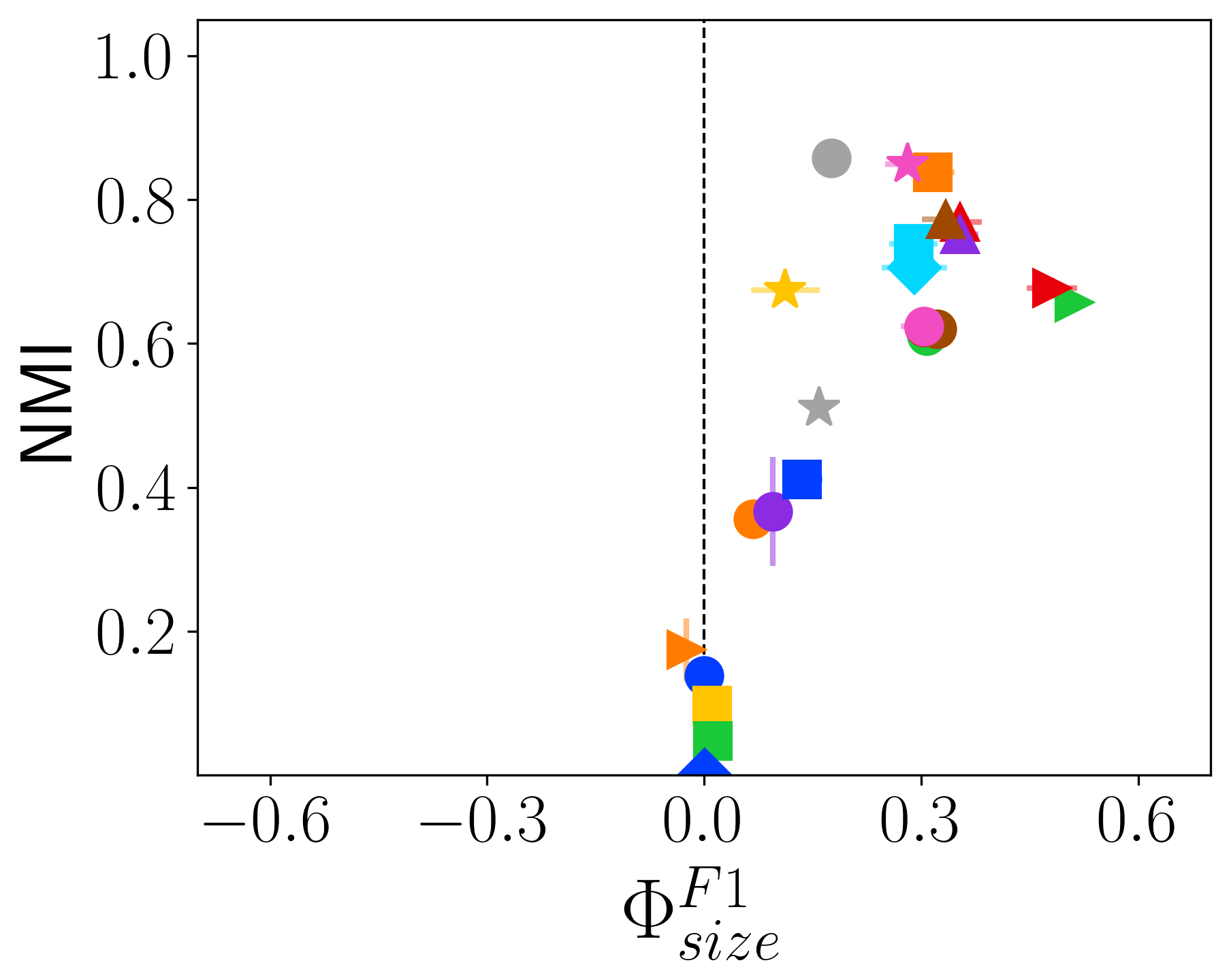}\quad
\includegraphics[width=0.31\textwidth]{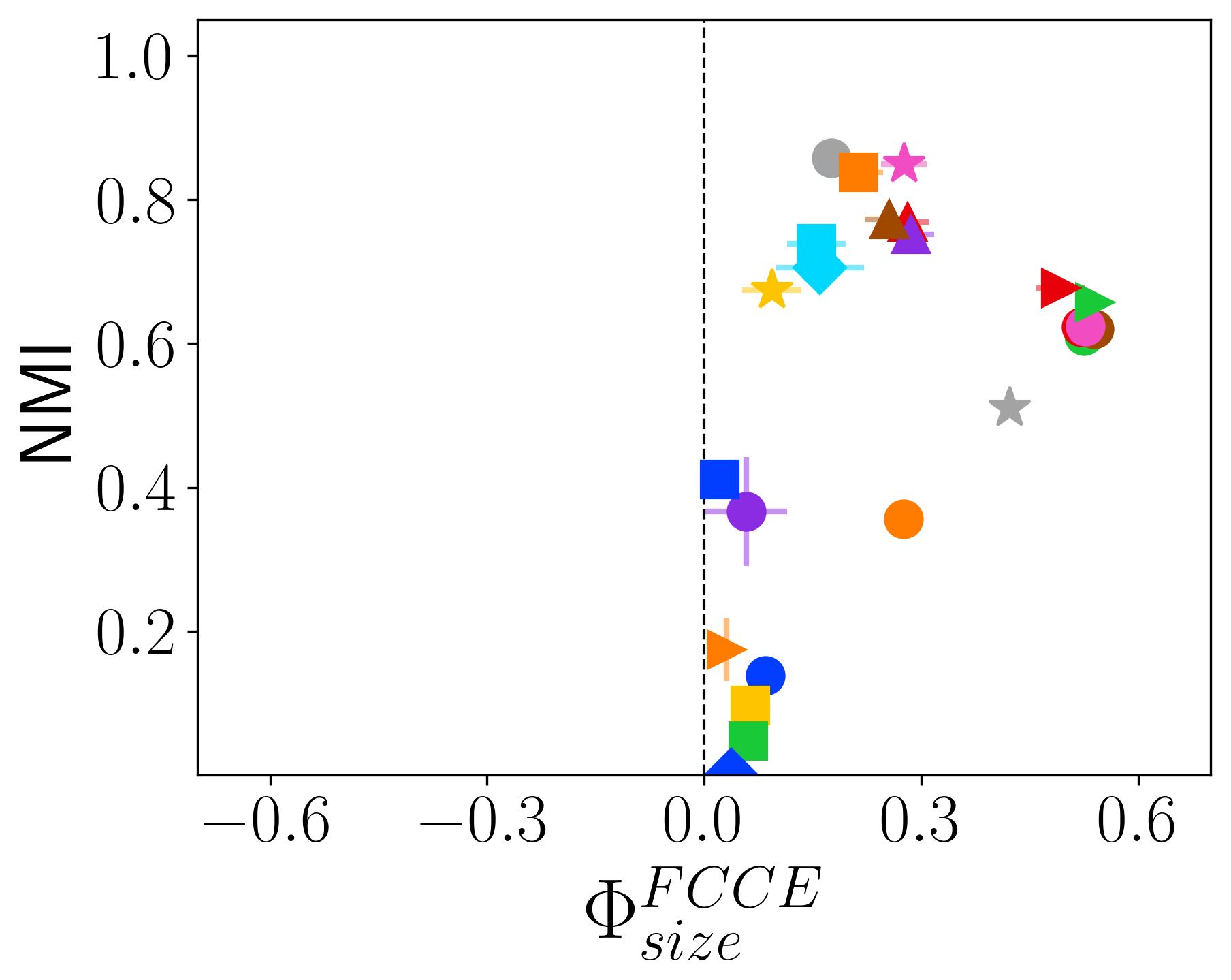}
\end{minipage}
\caption{NMI vs. fairness of community detection methods with respect to community size for LFR networks of 10,000 nodes having different $\mu$ values.}\label{phi_vs_size} 
\end{figure}

We first study how well different community detection methods identify communities of different sizes. In Fig. \ref{phi_vs_size}, we plot NMI versus three group fairness ($\Phi^{FCCN}_{size}$, $\Phi^{F1}_{size}$, and $\Phi^{FCCE}_{size}$) for LFR networks having mixing parameters $\mu =0.2,$ $0.4,$ and $0.6$. To perform our experiments, we create 10 LFR networks for each setting and apply CD methods (stochastic CD methods are executed ten times) to compute the average and standard deviation.

For $\mu=0.2$, all CD methods, except EM, Paris, SBM-nested, and RSC-SSE, favor large-size communities for all fairness metrics $F*$. The methods which have high fairness and identify good quality communities include RSC-K, RSC-V, RSC-SSE, Infomap, Fluid, Walktrap, and Significance (NMI $\sim$ 1). SBM-Nested identifies smaller communities (negative $\Phi^{F*}_{size}$) well as this is a hierarchical method, and at lower levels, it perfectly discovers smaller communities. When $\mu$ is increased to 0.4, the SBM-Nested method still favors small communities. However, when communities are highly connected with each other ($\mu=0.6$), no method identifies small groups well. All methods identify larger communities better, and the fair methods have very low NMI as they identify communities of all sizes equally bad. One important point to note is that across all networks, there is no correlation between the fairness and performance of CD methods. 

\begin{figure}[t]
\centering
\begin{subfigure}[b]{0.99\textwidth}            
    \includegraphics[width=\textwidth]{legend_ncol6.png}
\end{subfigure}\\
\begin{subfigure}[c]{0.05\textwidth}
\caption*{\rotatebox{90}{$\mu=0.2$}}
\end{subfigure}%
\begin{minipage}[c]{0.95\textwidth}
\includegraphics[width=0.31\textwidth]{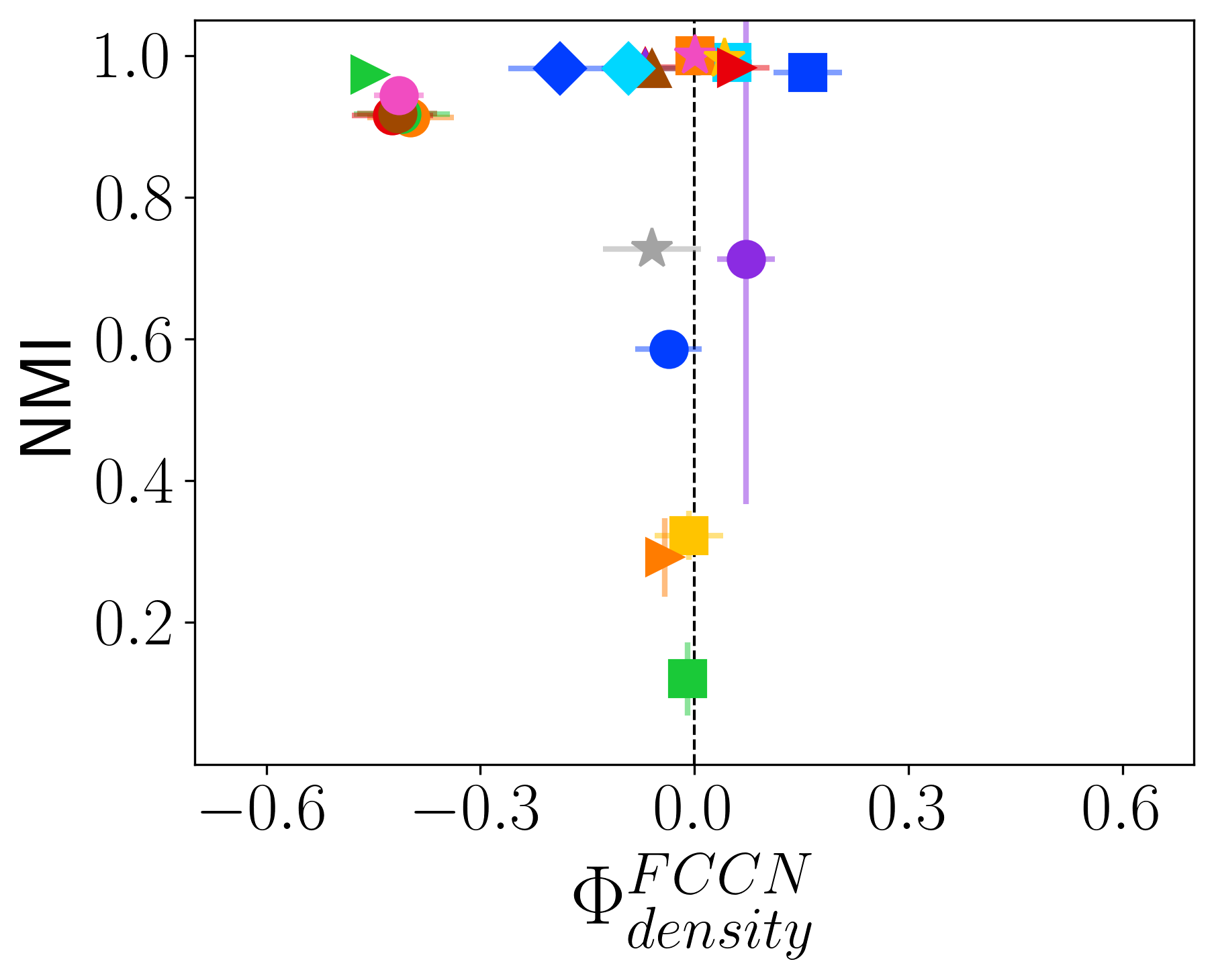}\quad
\includegraphics[width=0.31\textwidth]{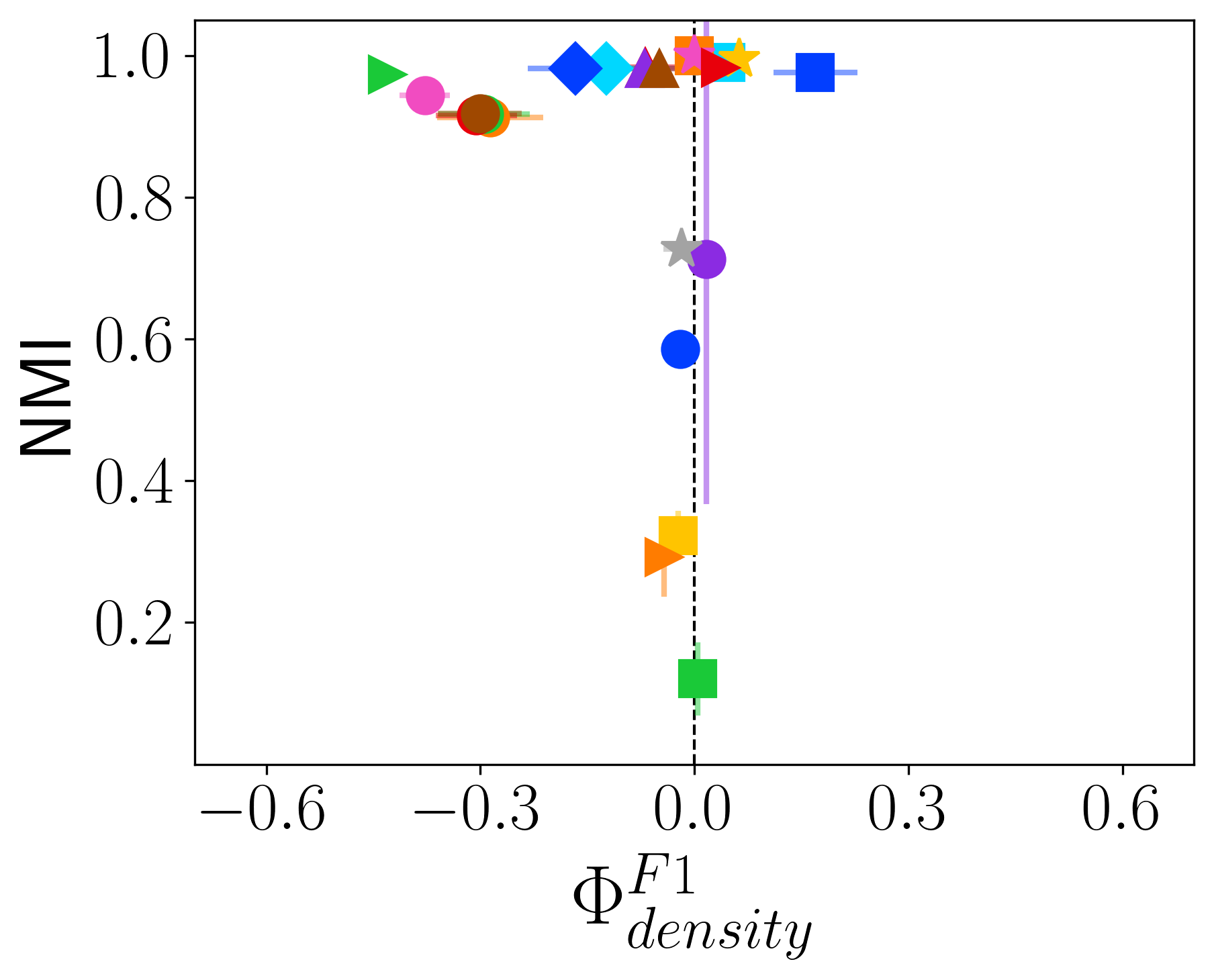}\quad
\includegraphics[width=0.31\textwidth]{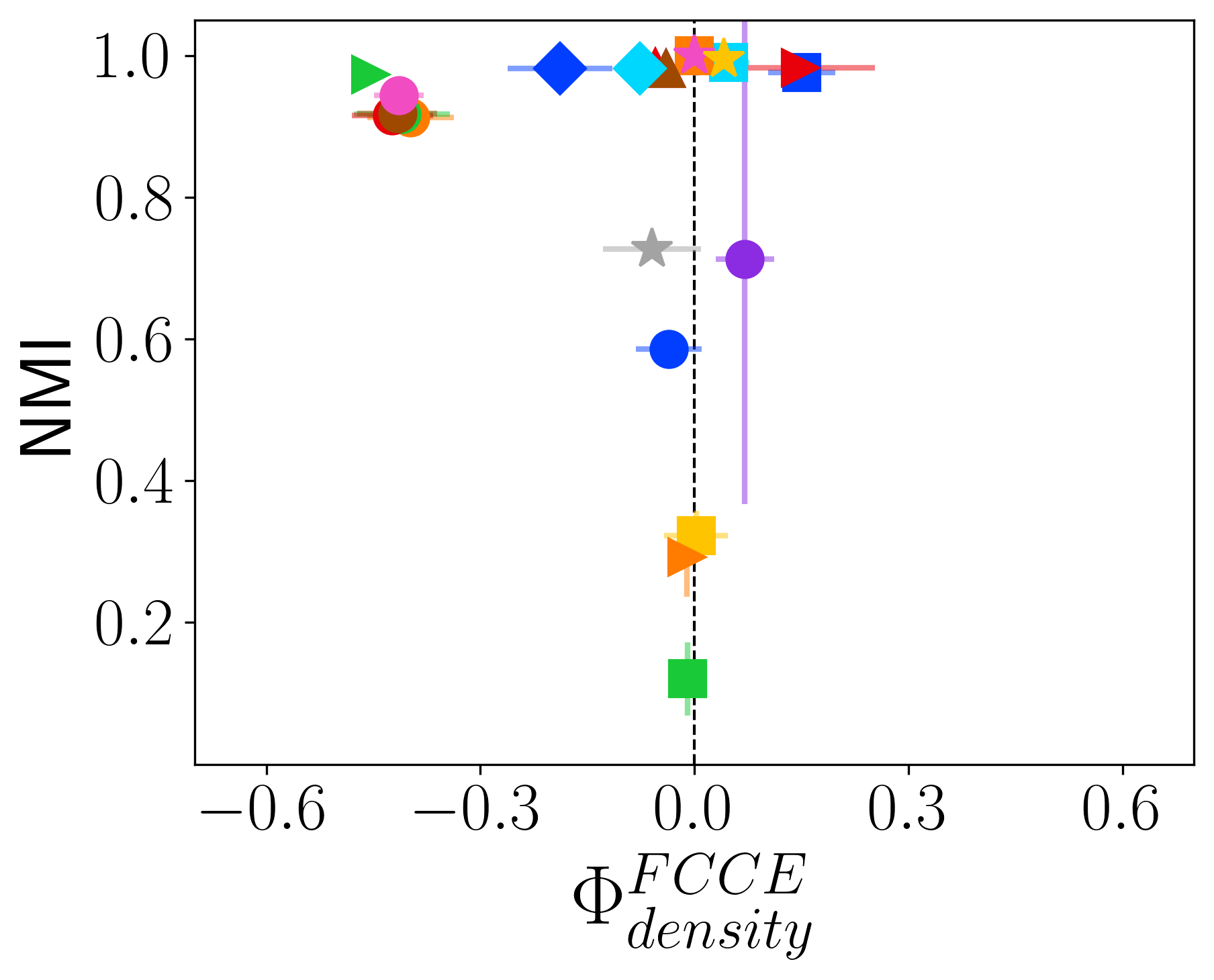}
\end{minipage}
\\
\begin{subfigure}[c]{0.05\textwidth}
\caption*{\rotatebox{90}{$\mu=0.4$}}
\end{subfigure}%
\begin{minipage}[c]{0.95\textwidth}
\includegraphics[width=0.31\textwidth]{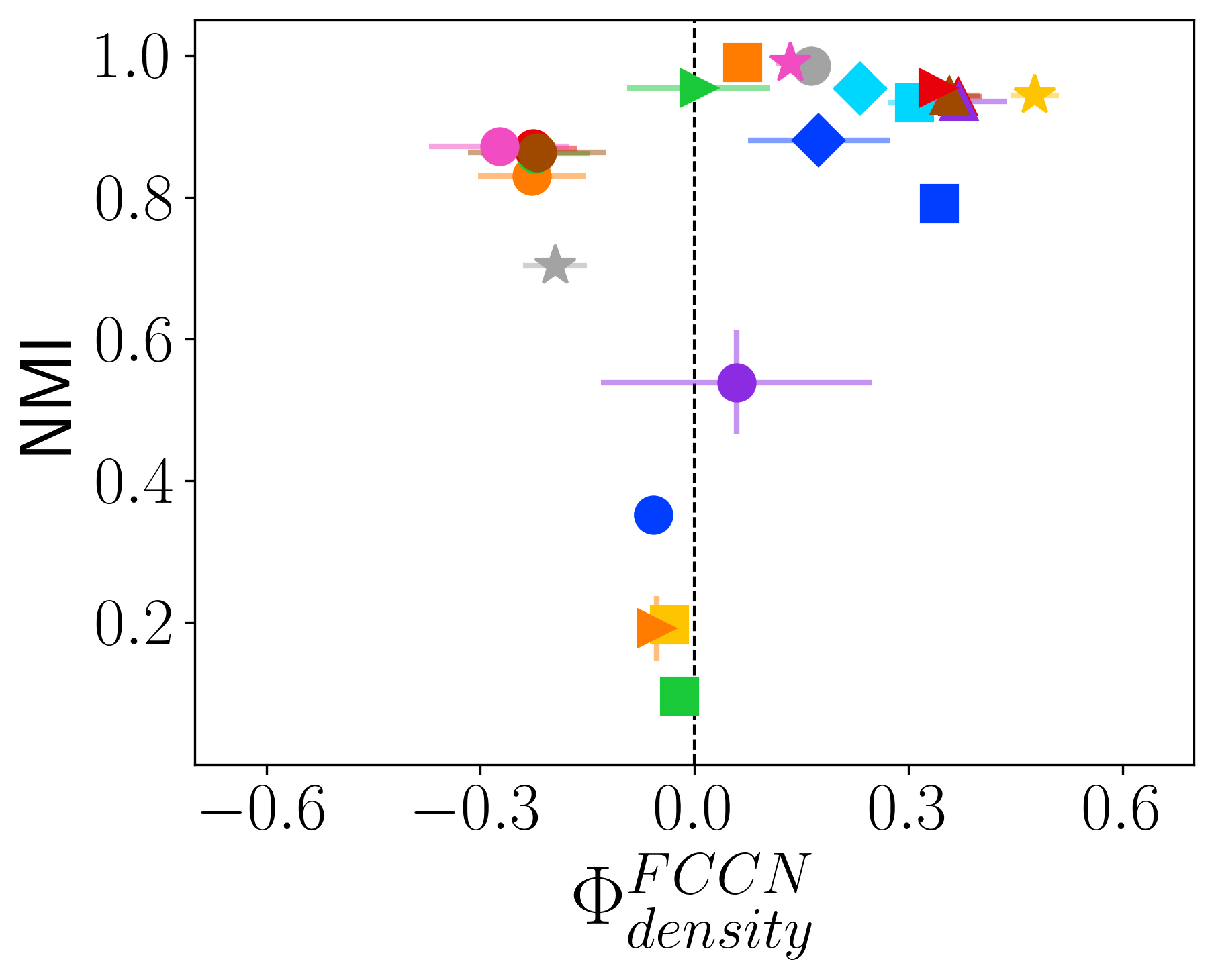}\quad
\includegraphics[width=0.31\textwidth]{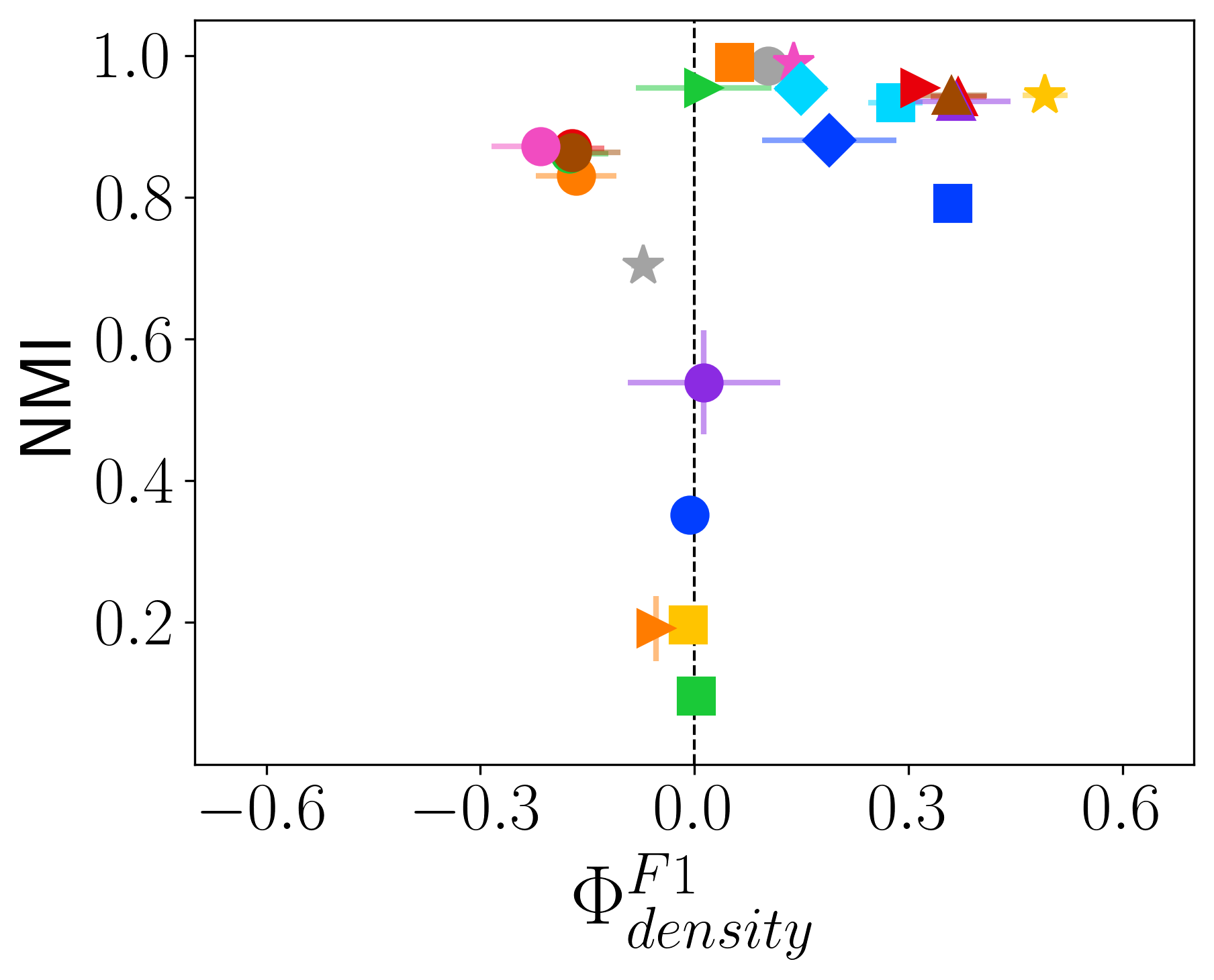}\quad
\includegraphics[width=0.31\textwidth]{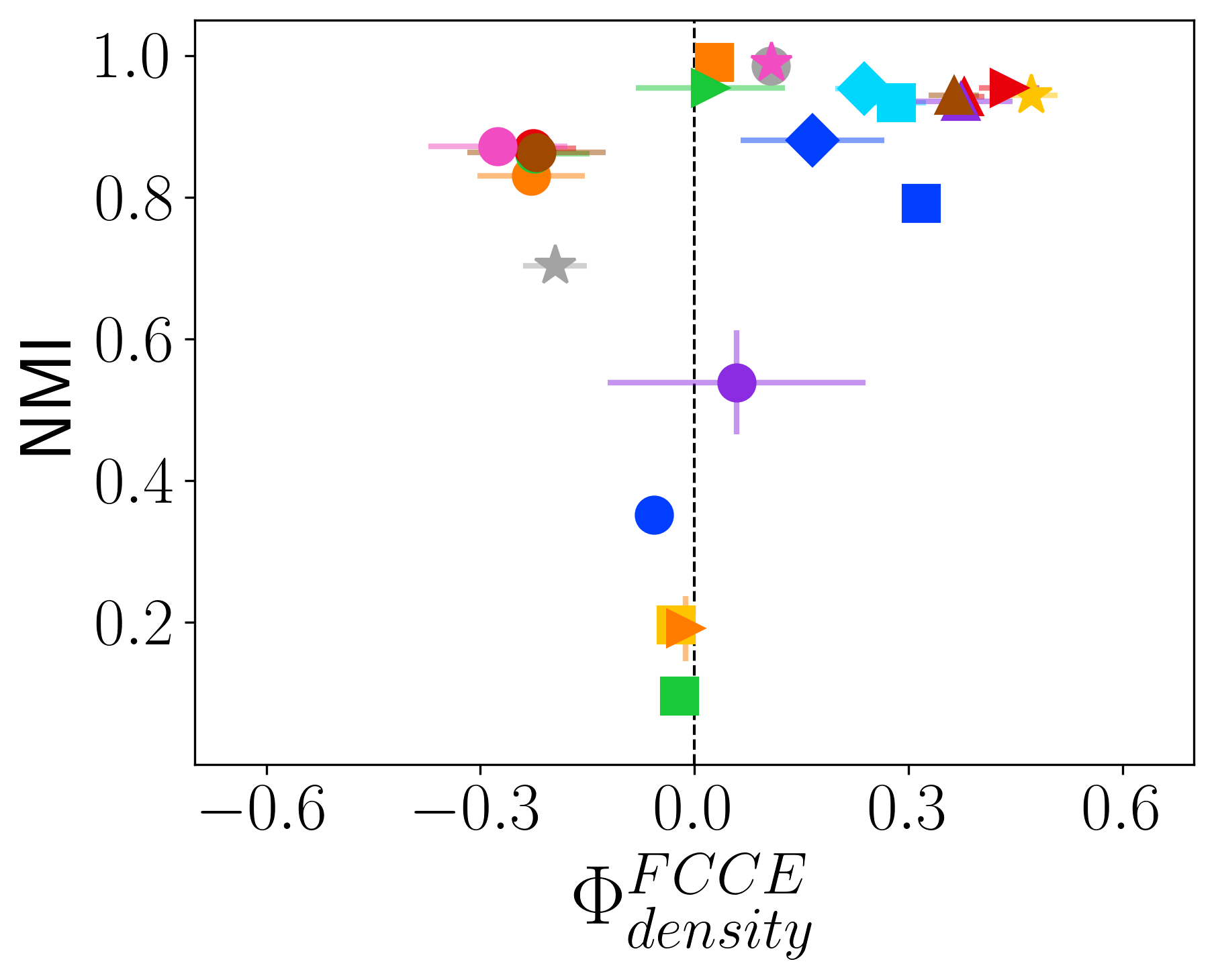}
\end{minipage}
\\
\begin{subfigure}[c]{0.05\textwidth}
\caption*{\rotatebox{90}{$\mu=0.6$}}
\end{subfigure}%
\begin{minipage}[c]{0.95\textwidth}
\includegraphics[width=0.31\textwidth]{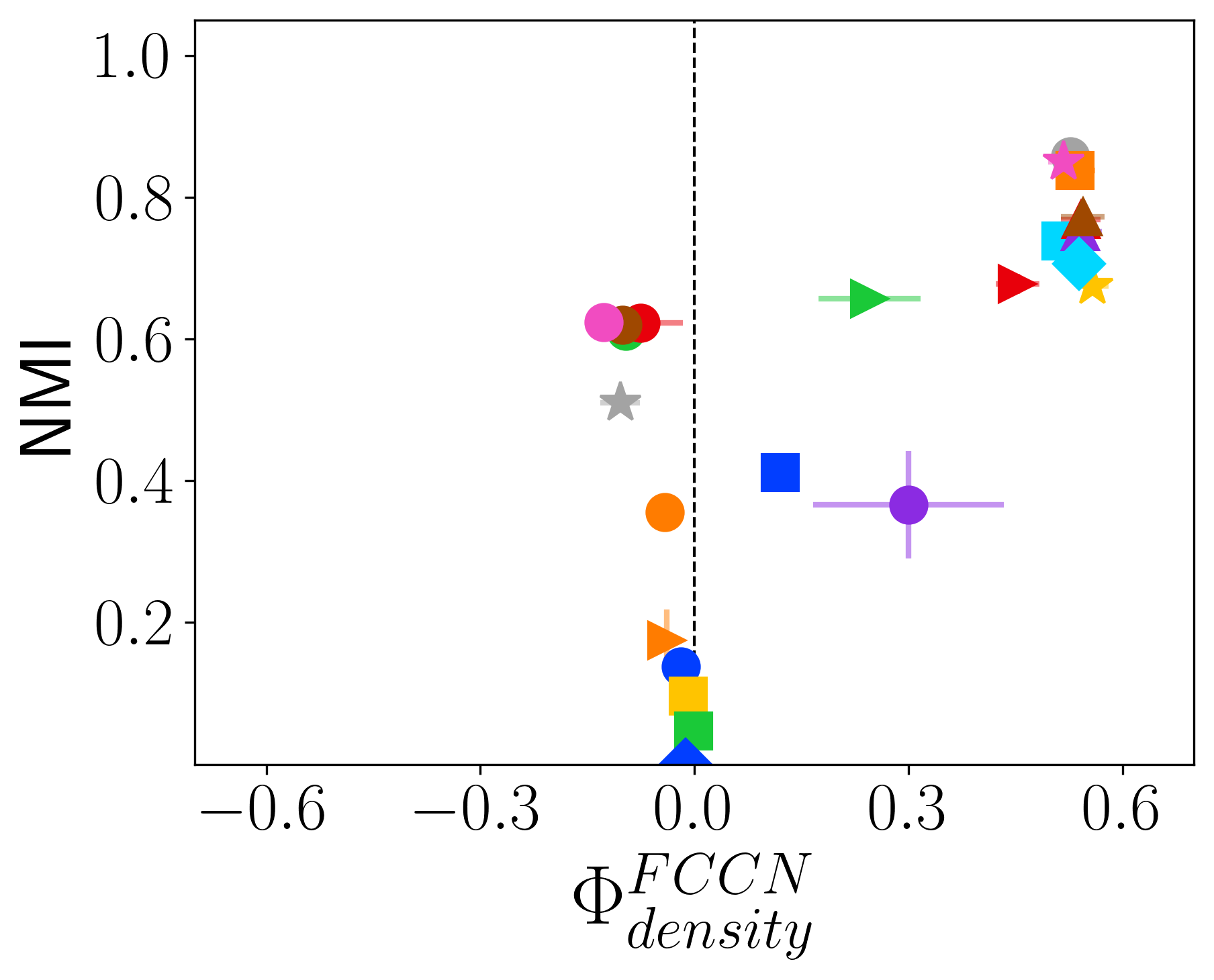}\quad
\includegraphics[width=0.31\textwidth]{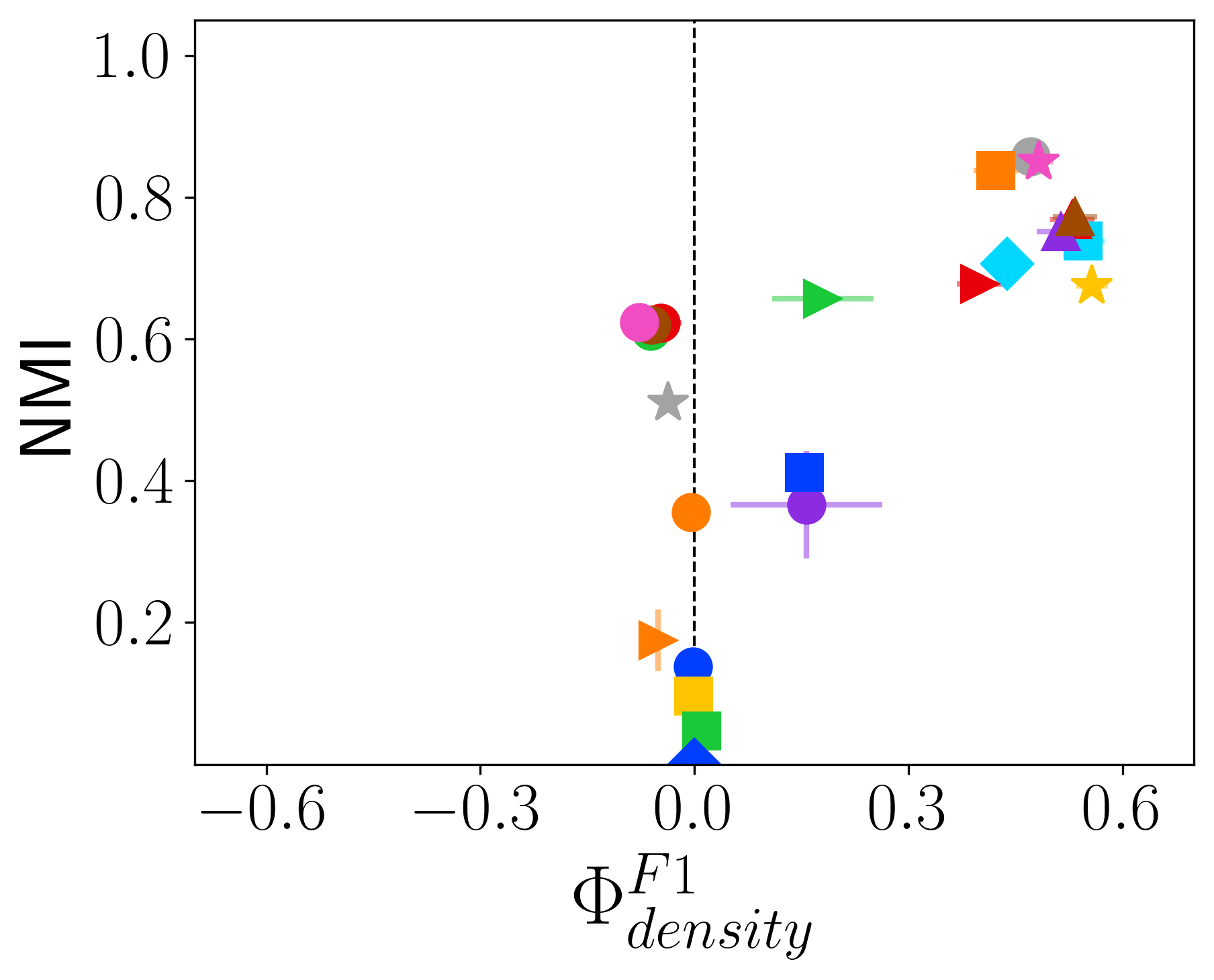}\quad
\includegraphics[width=0.31\textwidth]{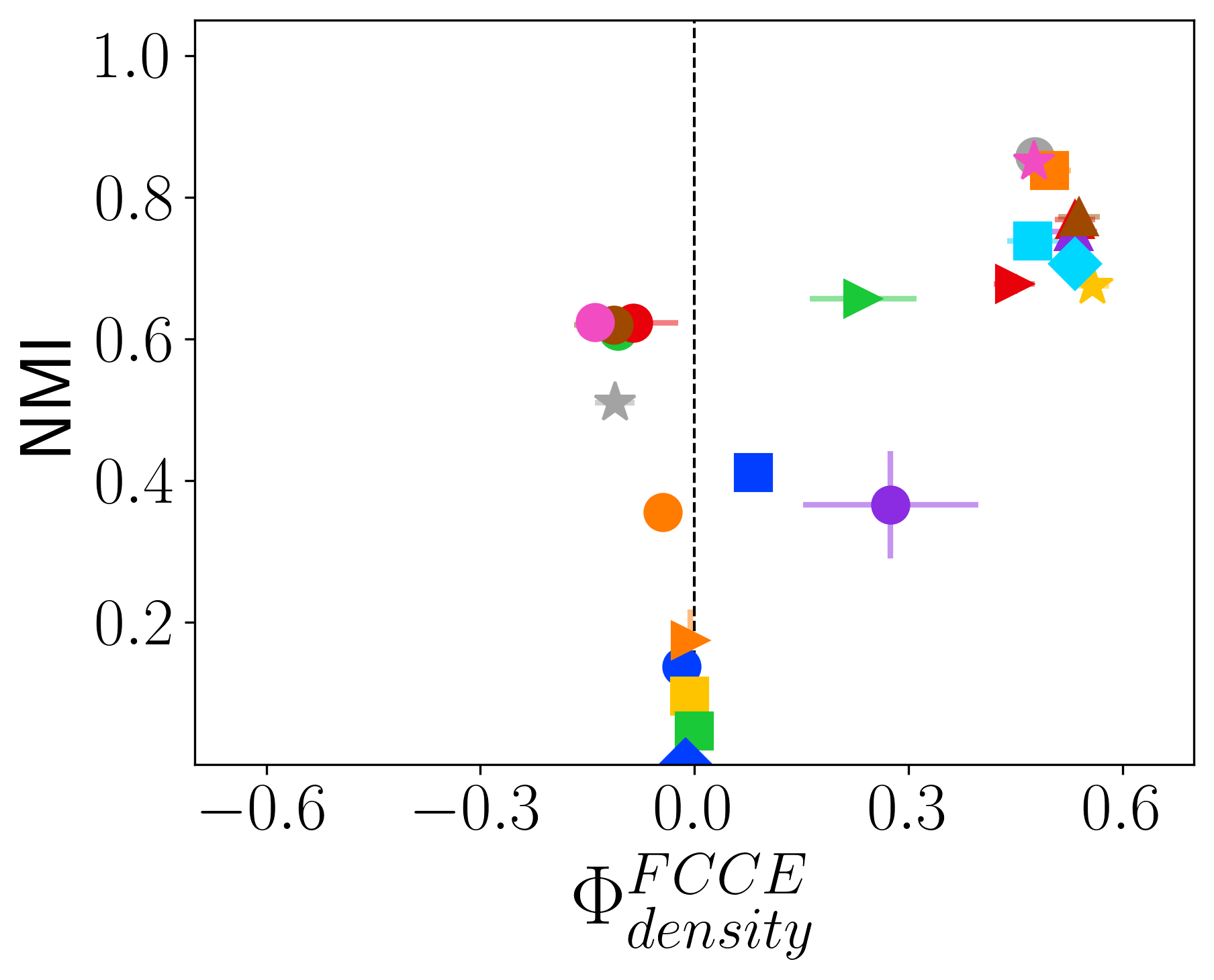}
\end{minipage}
\caption{NMI vs. fairness with respect to community densities on LFR networks.}\label{phi_vs_density} 
\end{figure}

\begin{figure}[t]
\centering
\begin{subfigure}[b]{0.98\textwidth}            
    \includegraphics[width=\textwidth]{legend_ncol6.png}
\end{subfigure}\\
\begin{subfigure}[c]{0.05\textwidth}
\caption*{\rotatebox{90}{$\mu=0.2$}}
\end{subfigure}%
\begin{minipage}[c]{0.95\textwidth}
\includegraphics[width=0.31\textwidth]{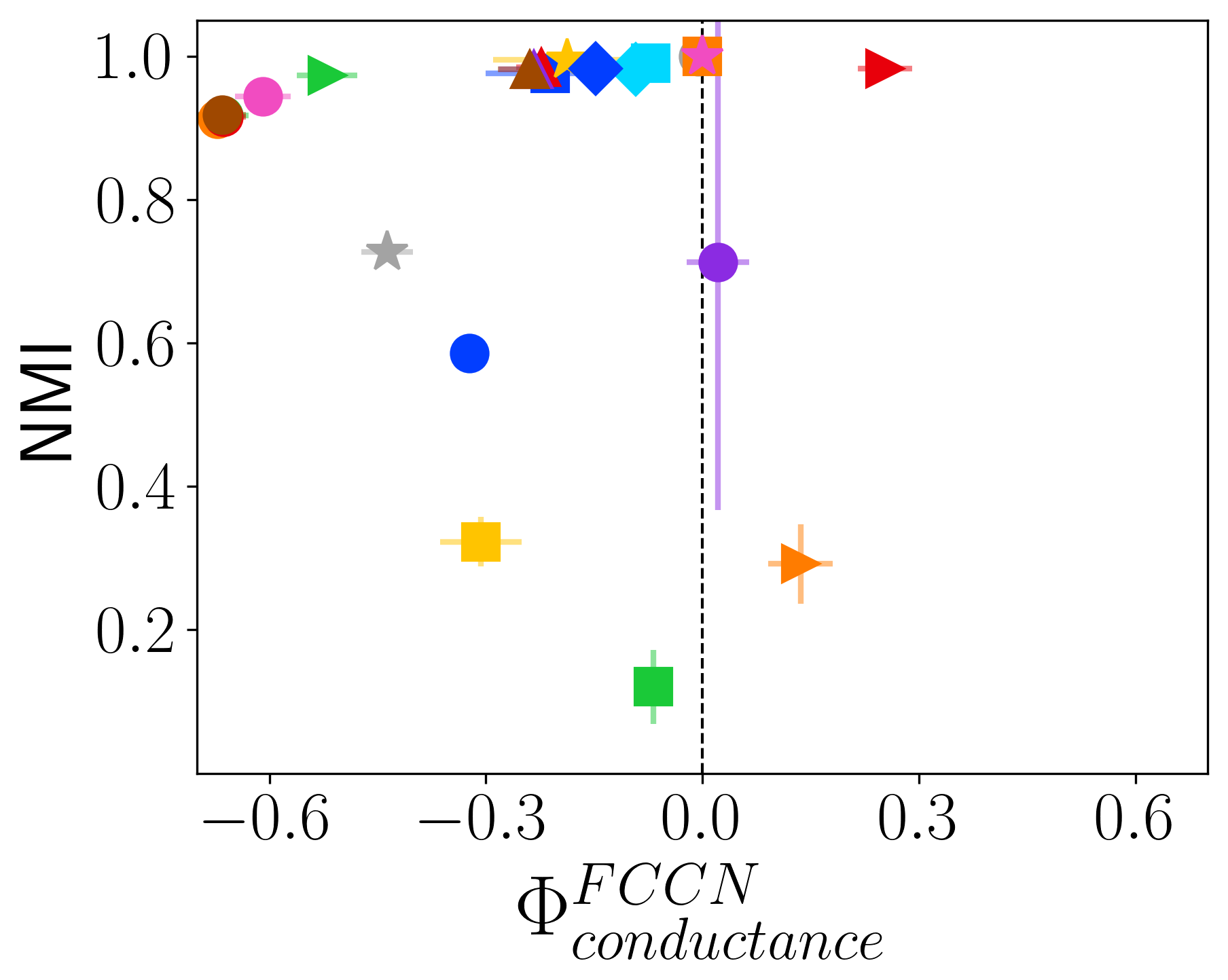}\quad
\includegraphics[width=0.31\textwidth]{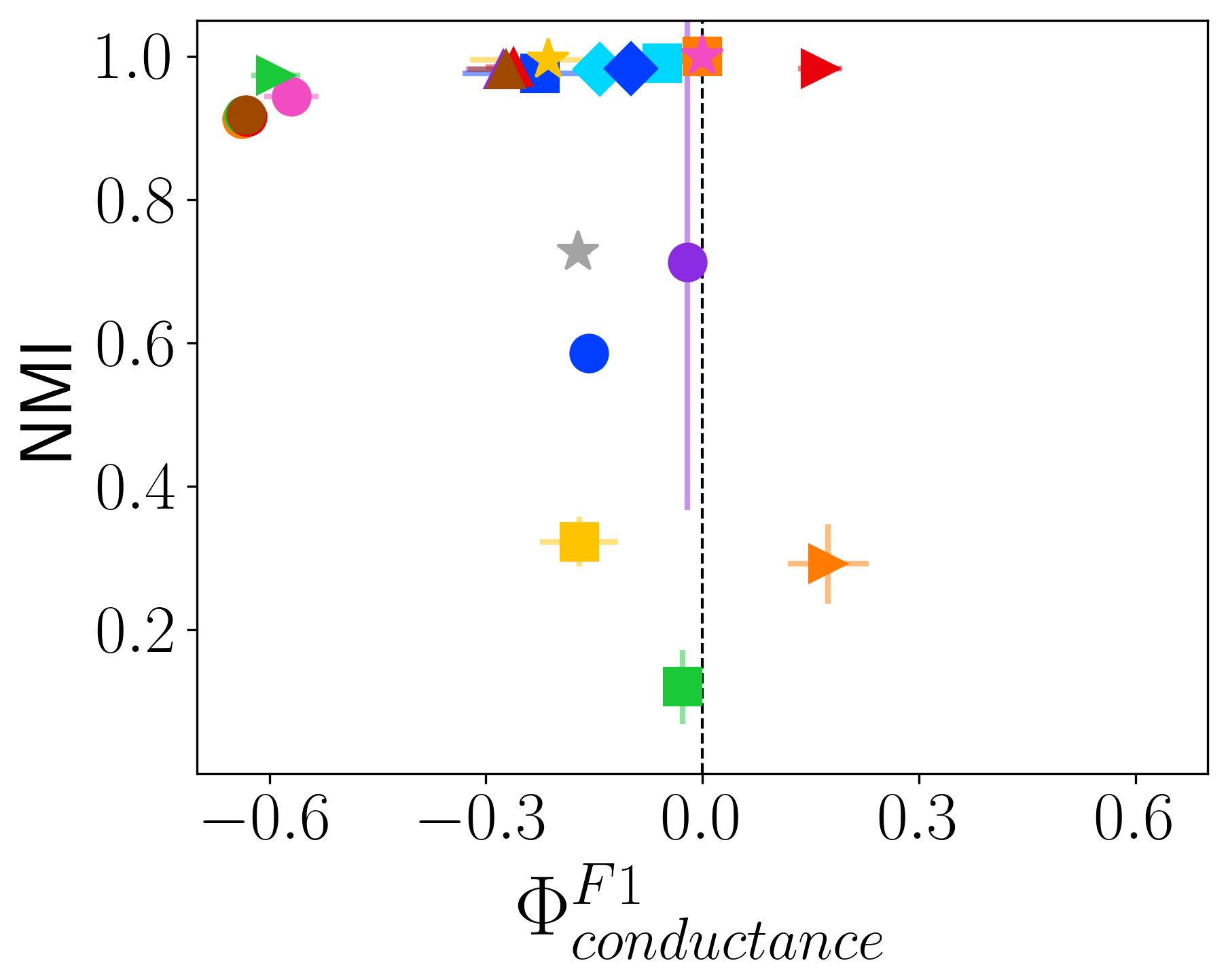}\quad
\includegraphics[width=0.31\textwidth]{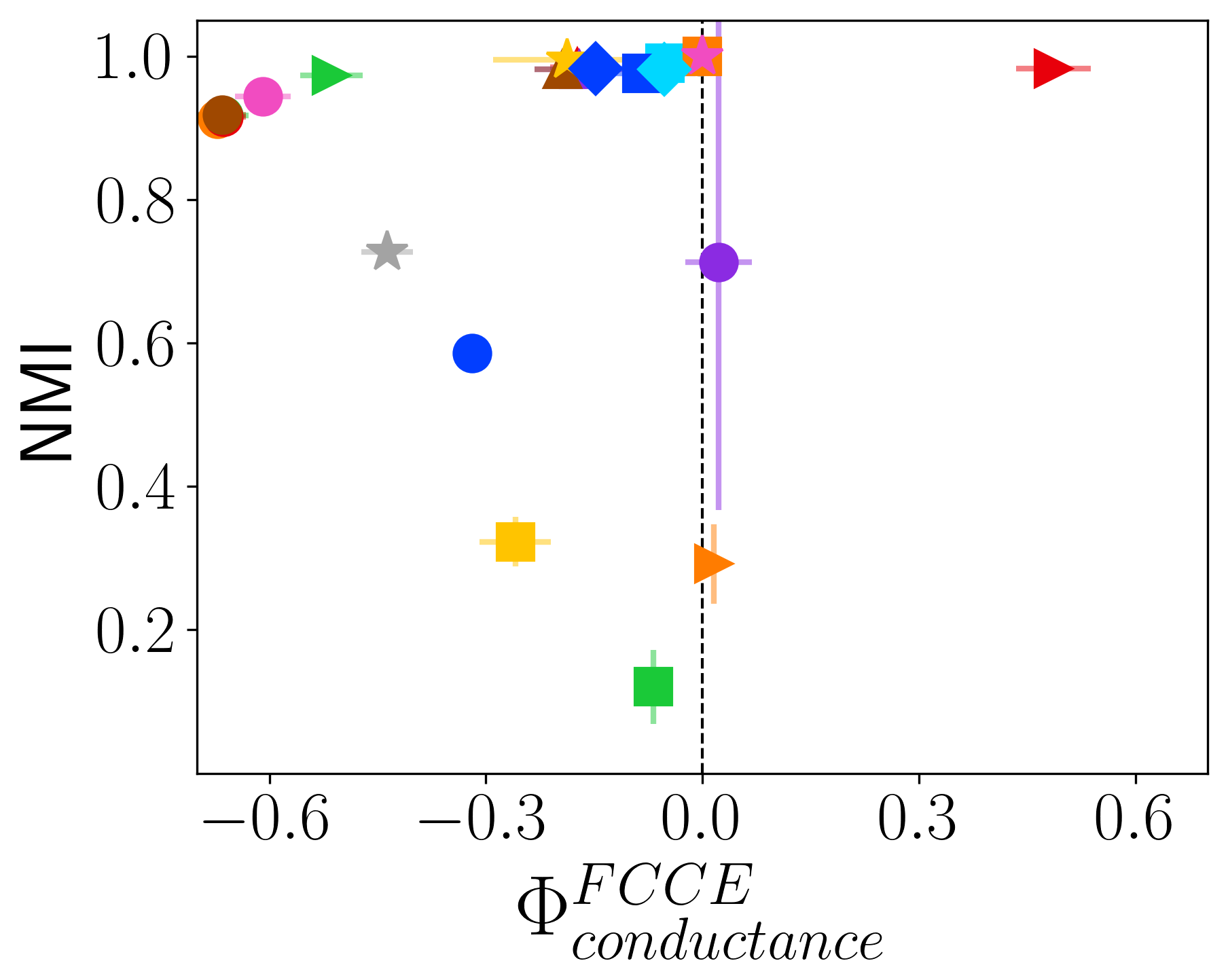}
\end{minipage}
\\
\begin{subfigure}[c]{0.05\textwidth}
\caption*{\rotatebox{90}{$\mu=0.4$}}
\end{subfigure}%
\begin{minipage}[c]{0.95\textwidth}
\includegraphics[width=0.31\textwidth]{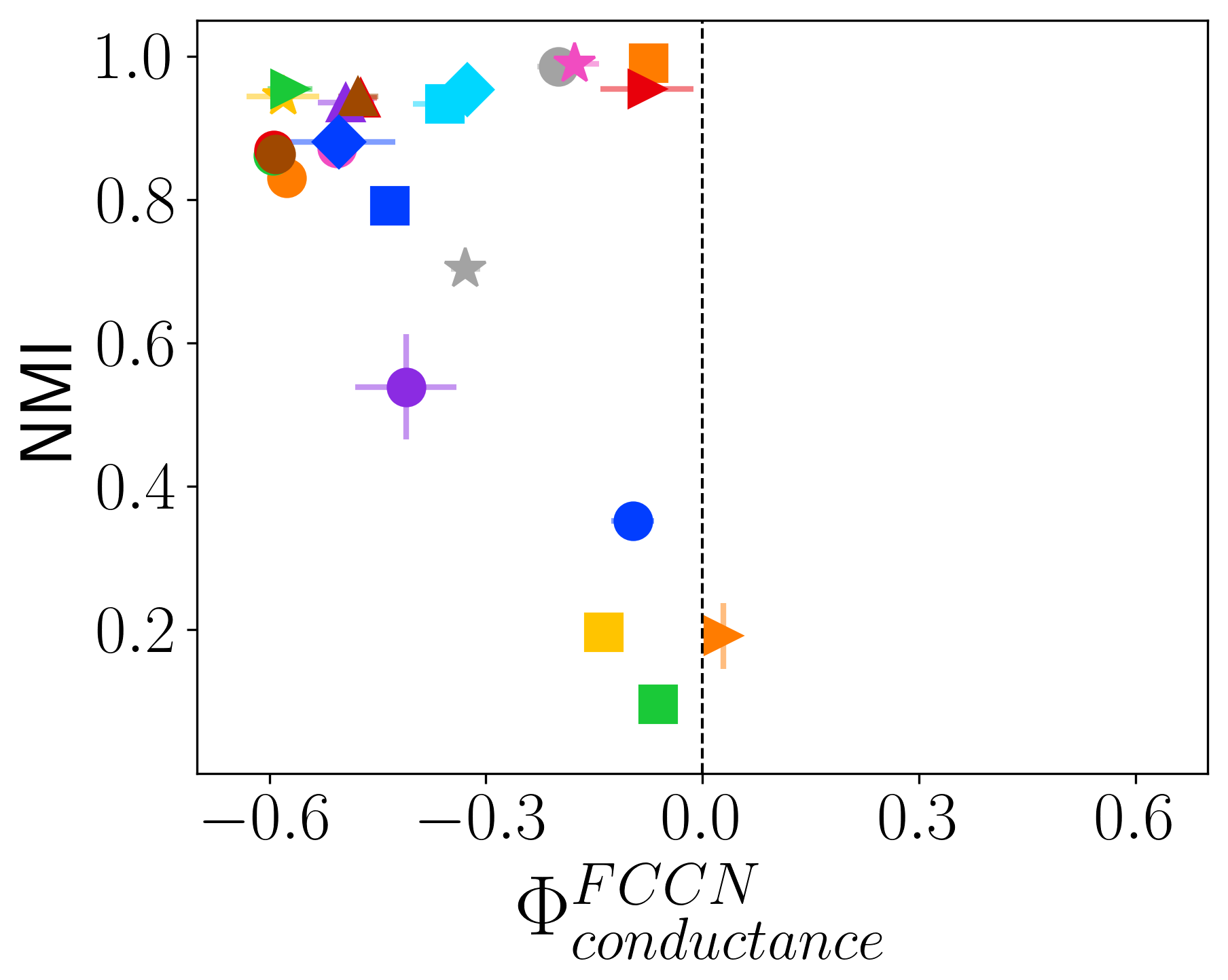}\quad
\includegraphics[width=0.31\textwidth]{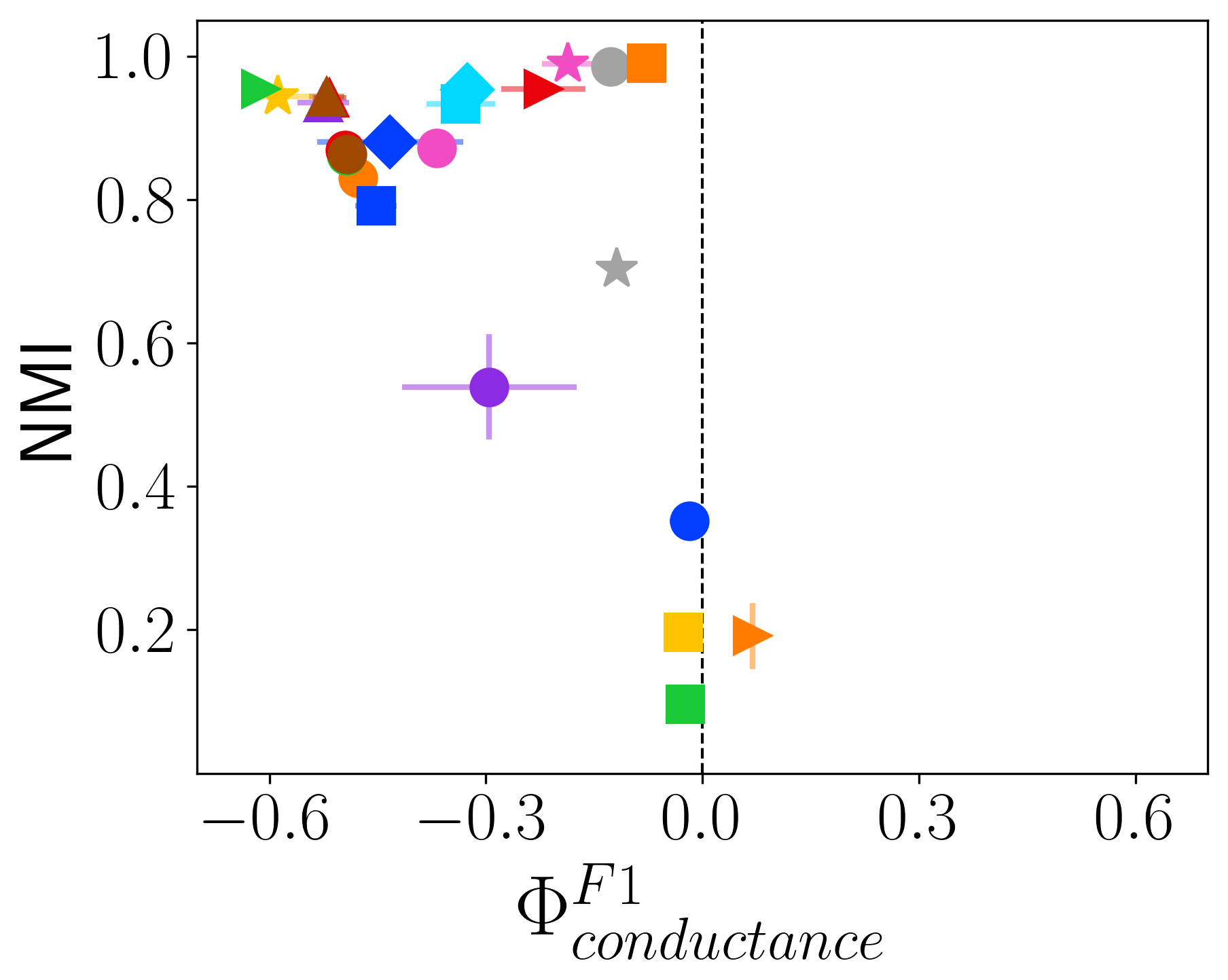}\quad
\includegraphics[width=0.31\textwidth]{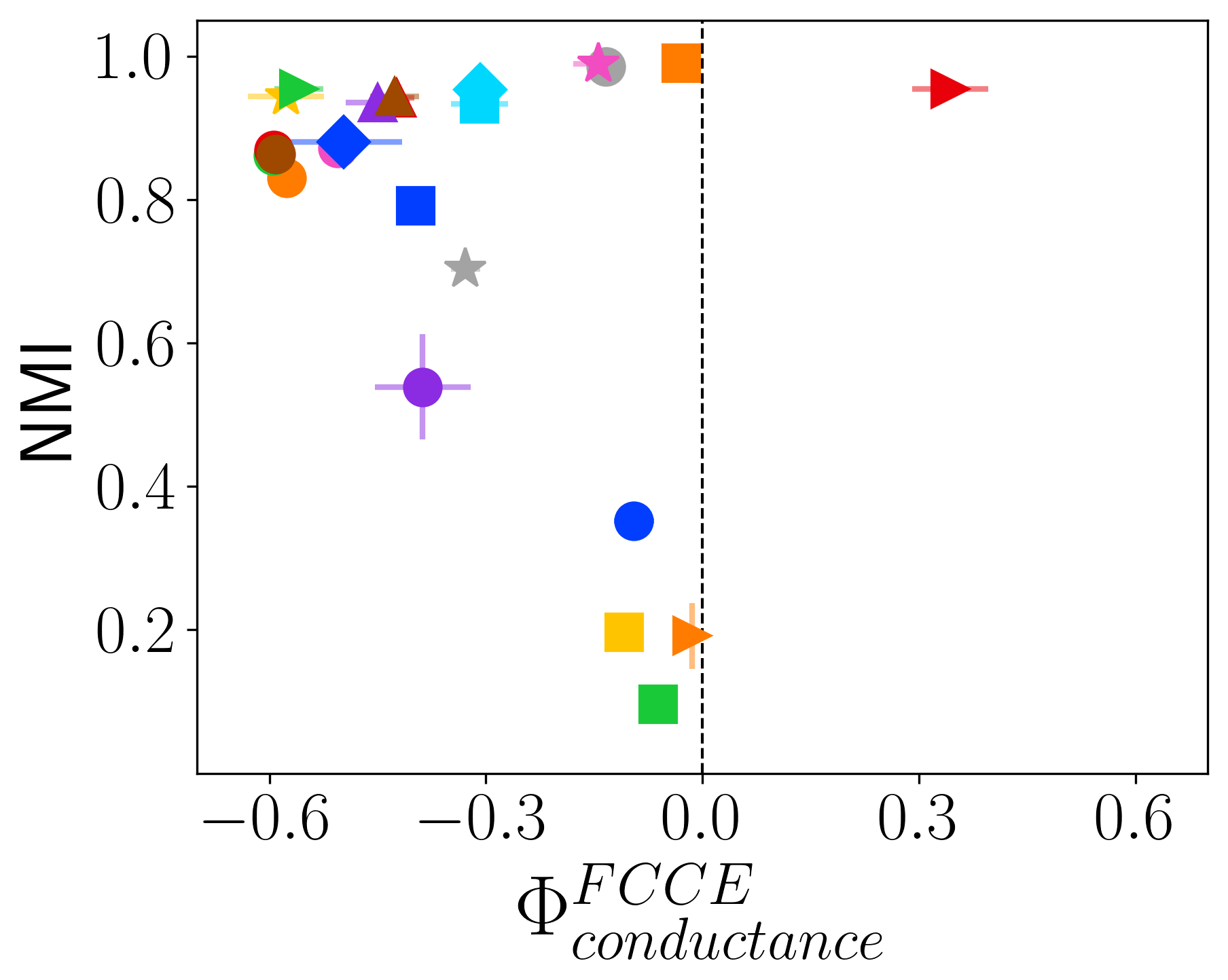}
\end{minipage}
\\
\begin{subfigure}[c]{0.05\textwidth}
\caption*{\rotatebox{90}{$\mu=0.6$}}
\end{subfigure}%
\begin{minipage}[c]{0.95\textwidth}
\includegraphics[width=0.31\textwidth]{result_plots_large/dens_mu6_mFCCN.png}\quad
\includegraphics[width=0.31\textwidth]{result_plots_large/dens_mu6_mF1.png}\quad
\includegraphics[width=0.31\textwidth]{result_plots_large/dens_mu6_mFCCE.png}
\end{minipage}
\caption{NMI vs. fairness with respect to conductance on LFR networks.}\label{phi_vs_conductance} 
\end{figure}

\subsection{Fairness Performance trade-off vs. Community Density} 

We further study the performance of CD methods for communities of different densities. In Fig.~\ref{phi_vs_density}, we observe that when the mixing parameter is low, CD methods identify sparse communities better than denser communities. However, this behavior changes as the mixing parameter is increased and the nodes have more inter-community edges. With $\mu=0.4,$ and $0.6$, Leiden, Louvain, RB-C, RB-ER, and Spinglass CD methods have similarly high NMI and negative $\Phi^{F*}_{size}$. All these methods predict fewer communities than the ground-truth, and predicted communities are mapped with low-density ground-truth communities. At $\mu=0.6$, methods with high NMI also predict dense communities better. For example, the Significance predicts a lot more communities than the ground-truth, and it identifies denser communities but splits up sparser communities in the prediction.

\subsection{Fairness Performance trade-off vs. Community Conductance}
Fig.~\ref{phi_vs_conductance} presents NMI vs. fairness with respect to community conductance. Most CD methods favor communities with lower conductance, meaning communities that have more separation. Across all fairness metrics, the bias increases with $\mu$, leading to lower $\Phi^{F*}_{conductance}$ values. With medium mixing ($\mu=0.4$), RSC-V, SBM-Nested, Infomap, and Significance are both fair and identify high-quality communities. At $\mu=0.6$, $\Phi^{F*}_{conductance}$ and NMI have linear relations, where fair methods have poor predictions and better predictions are highly biased.

\subsection{Fairness Metric versus Other Quality Evaluation Metrics}

\begin{figure}[t]
\centering
\begin{subfigure}[b]{0.98\textwidth}            
    \includegraphics[width=\textwidth]{legend_ncol6.png}
\end{subfigure}\\
\begin{subfigure}[c]{0.05\textwidth}
\caption*{\rotatebox{90}{ARI}}%
\end{subfigure}%
\begin{minipage}[c]{0.94\textwidth}
\includegraphics[width=0.31\textwidth]{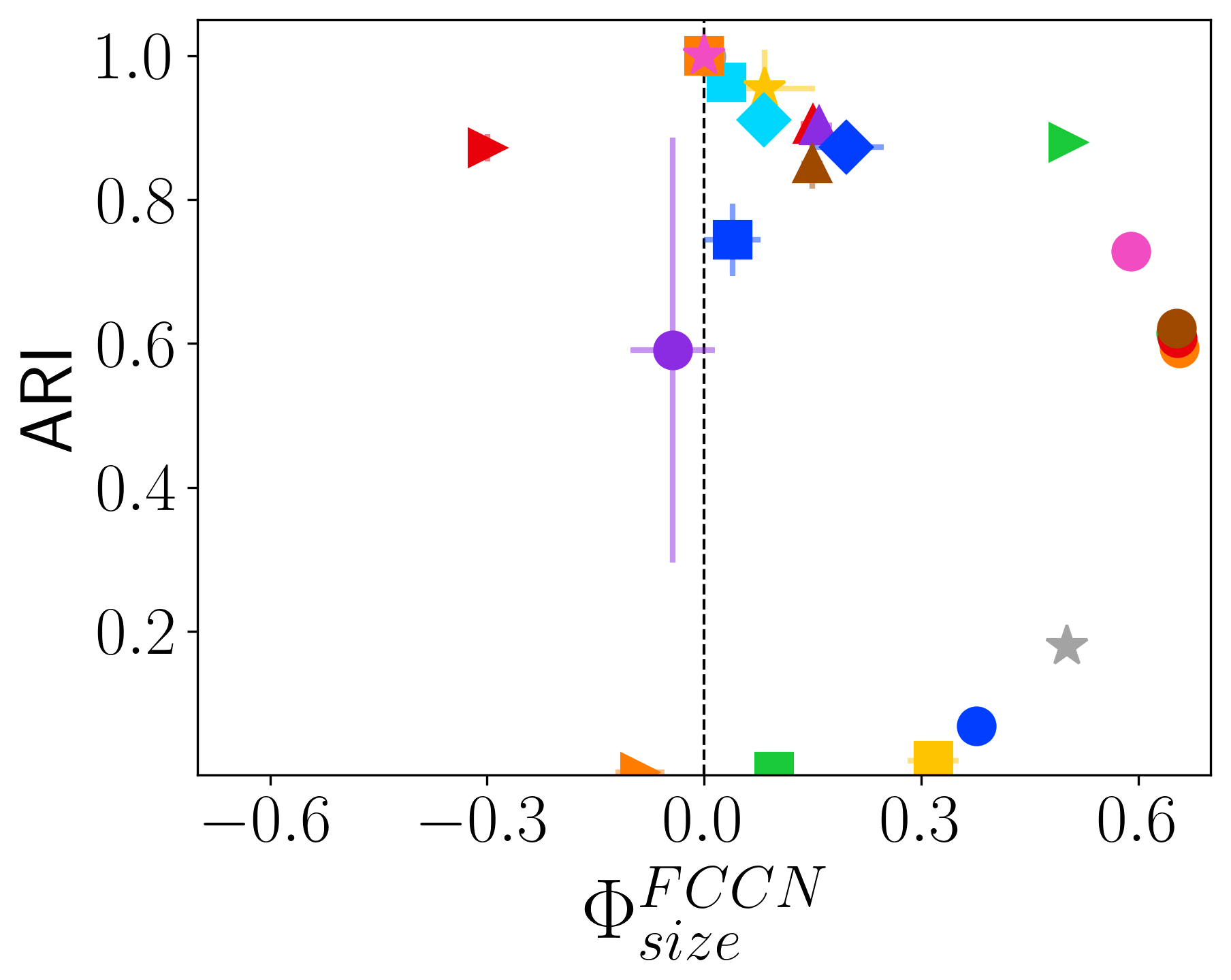}\quad
\includegraphics[width=0.31\textwidth]{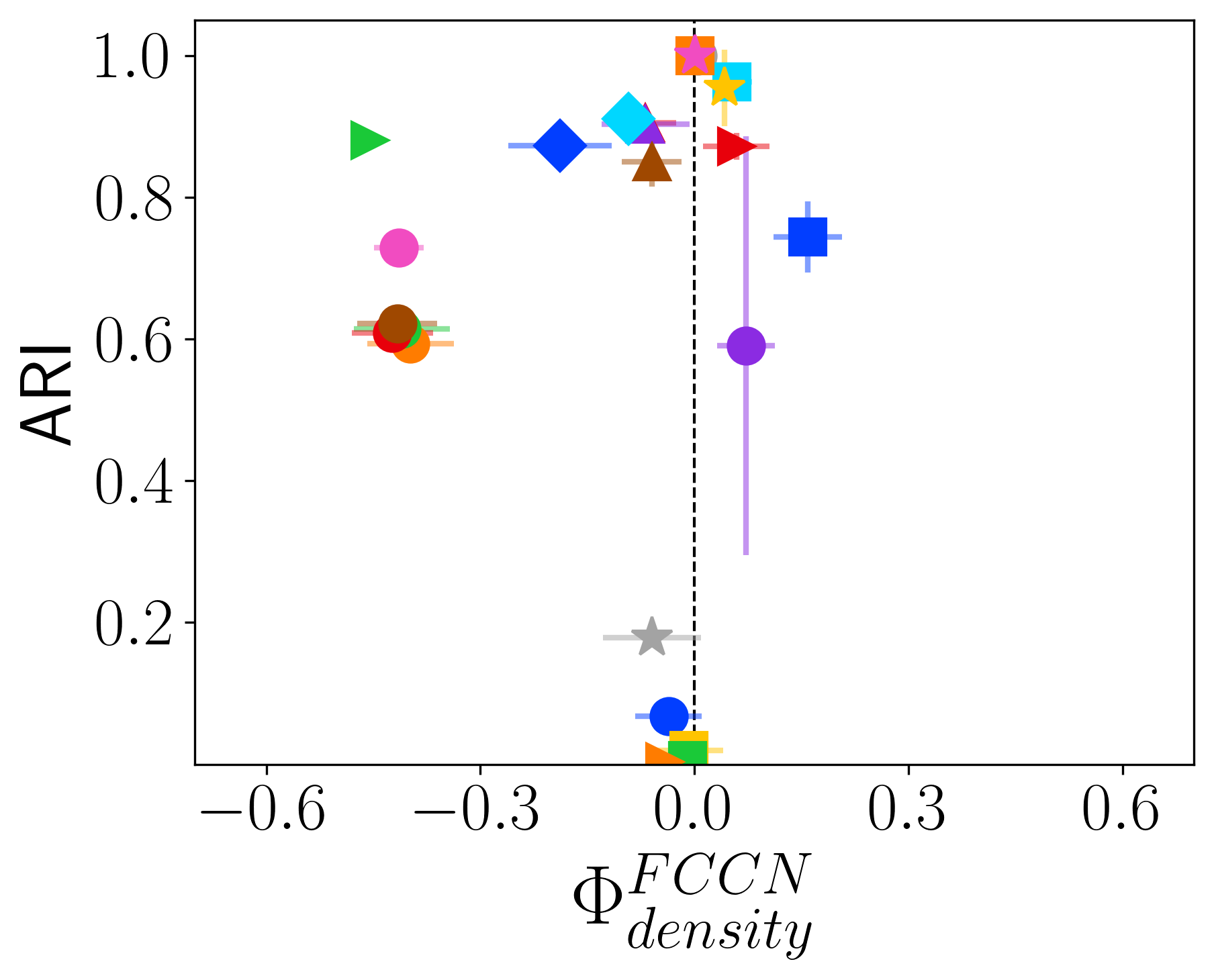}\quad
\includegraphics[width=0.31\textwidth]{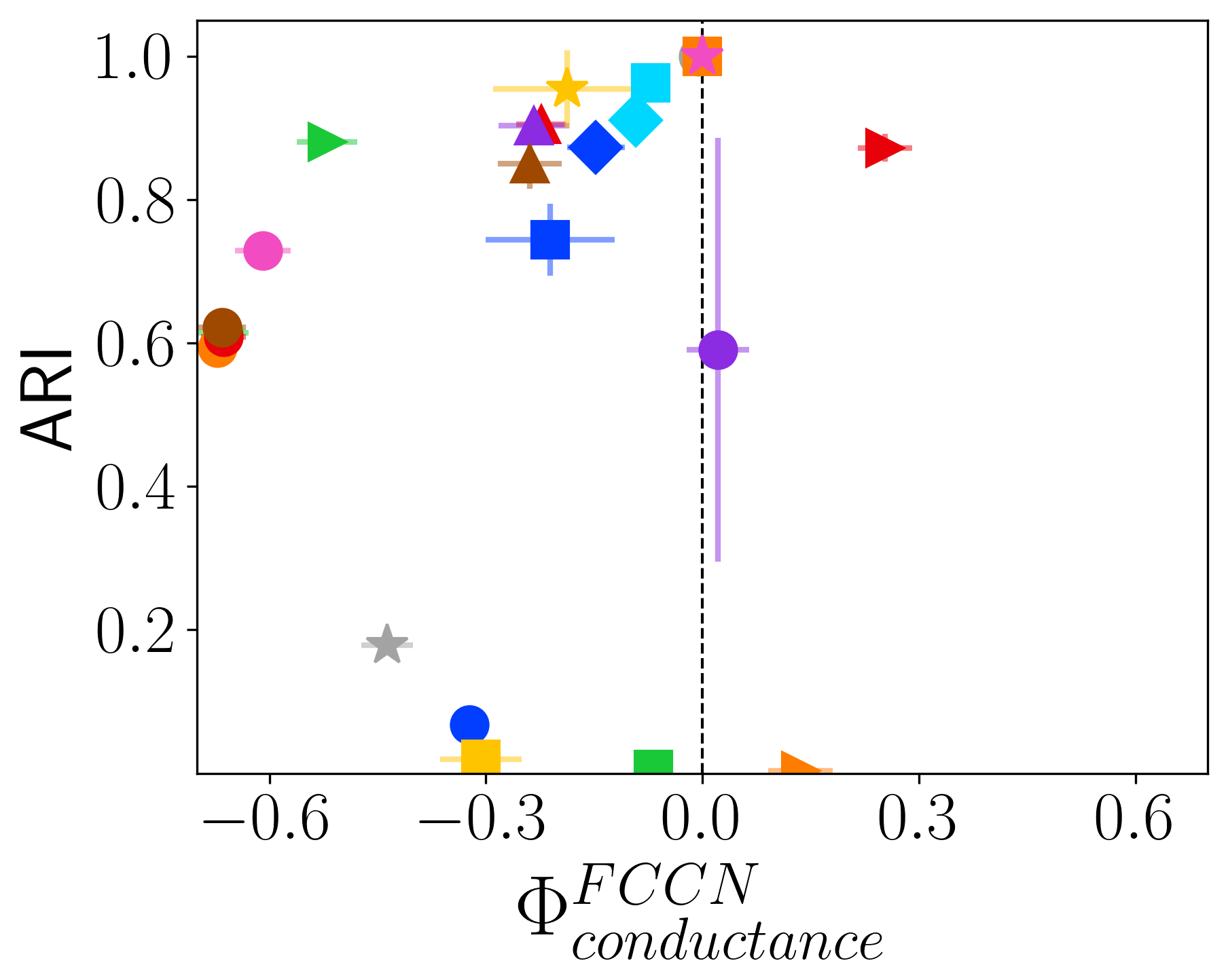}
\end{minipage}
\\
\begin{subfigure}[c]{0.05\textwidth}
\caption*{\rotatebox{90}{RMI}}
\end{subfigure}
\begin{minipage}[c]{0.94\textwidth}
\includegraphics[width=0.31\textwidth]{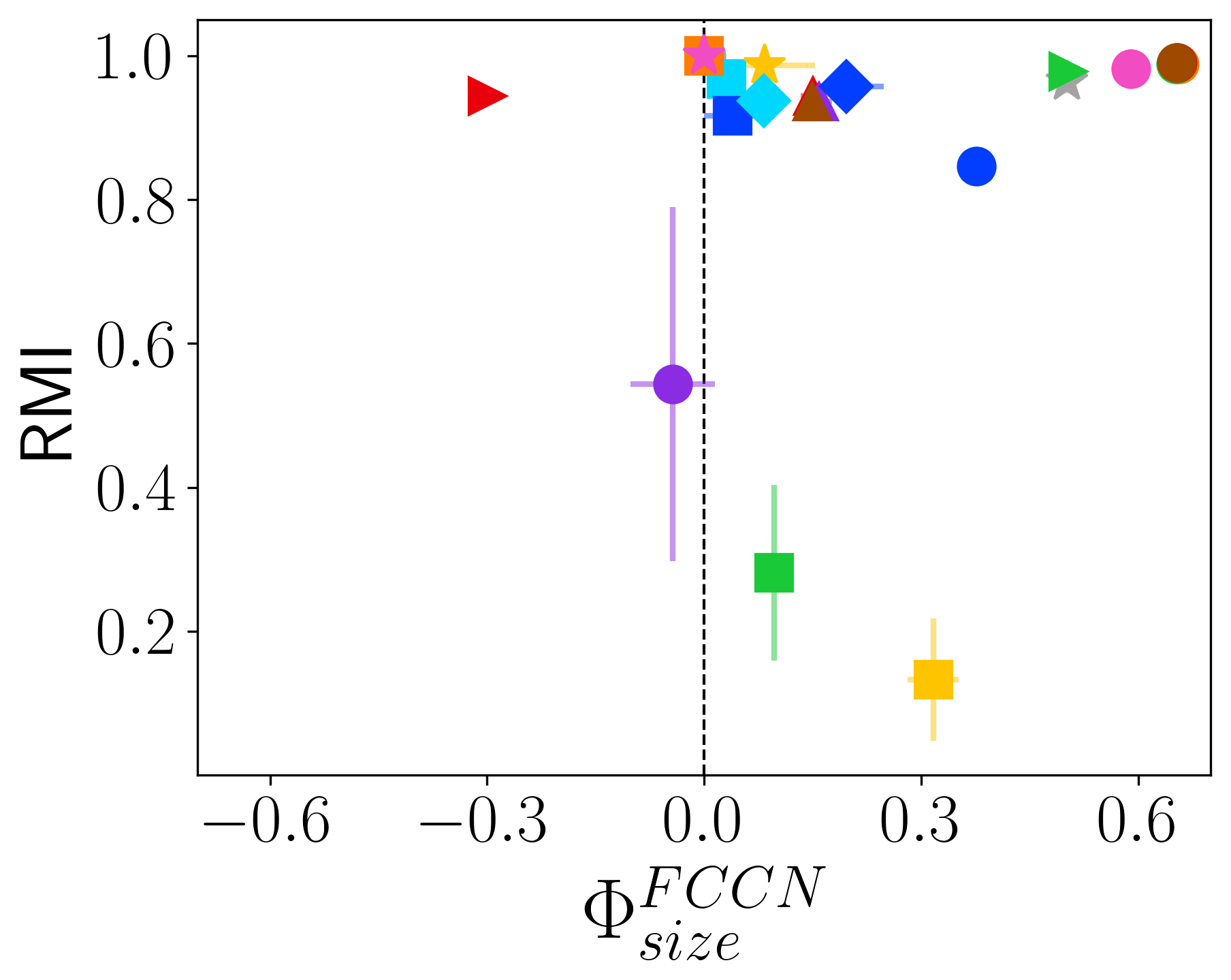}\quad
\includegraphics[width=0.31\textwidth]{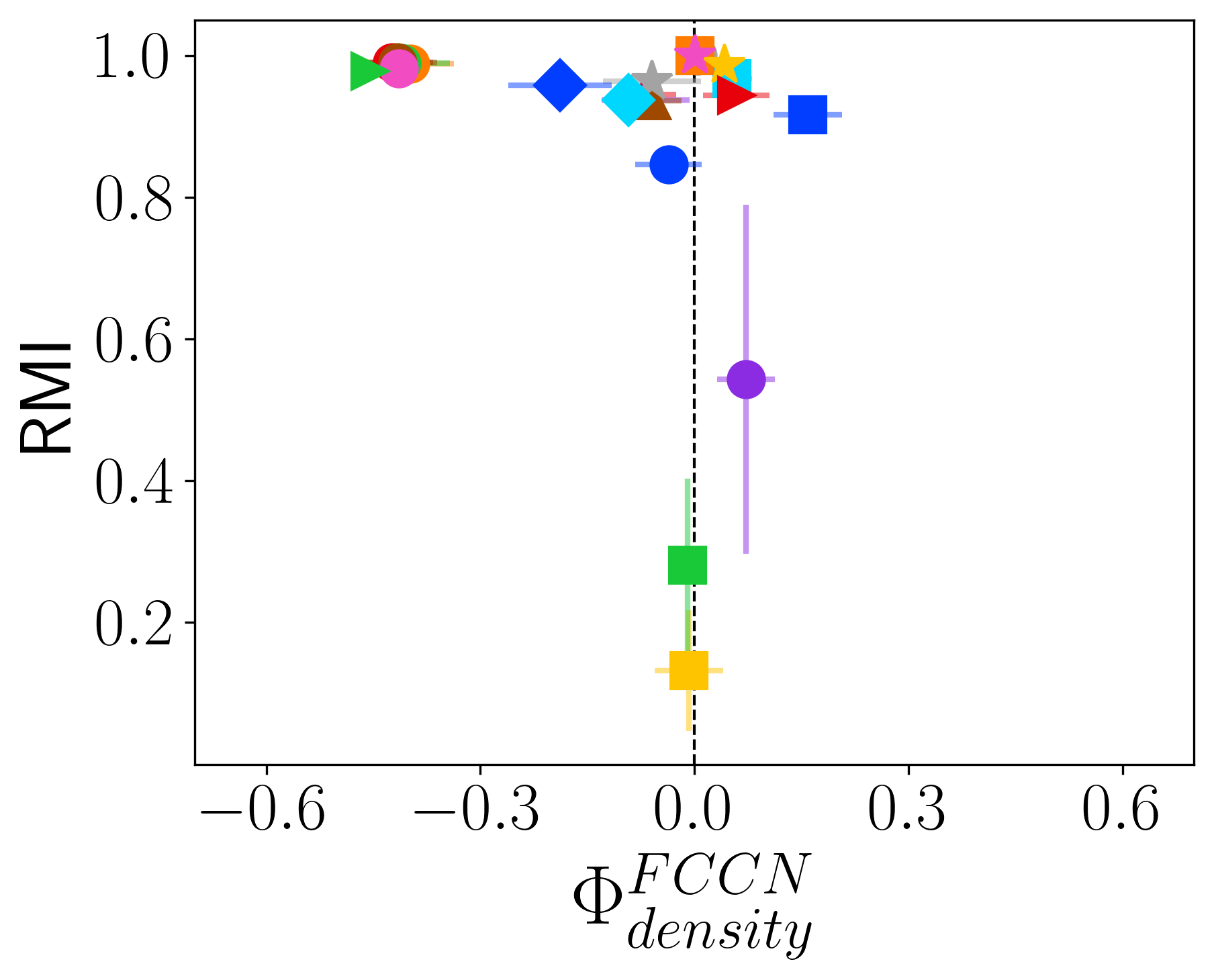}\quad
\includegraphics[width=0.31\textwidth]{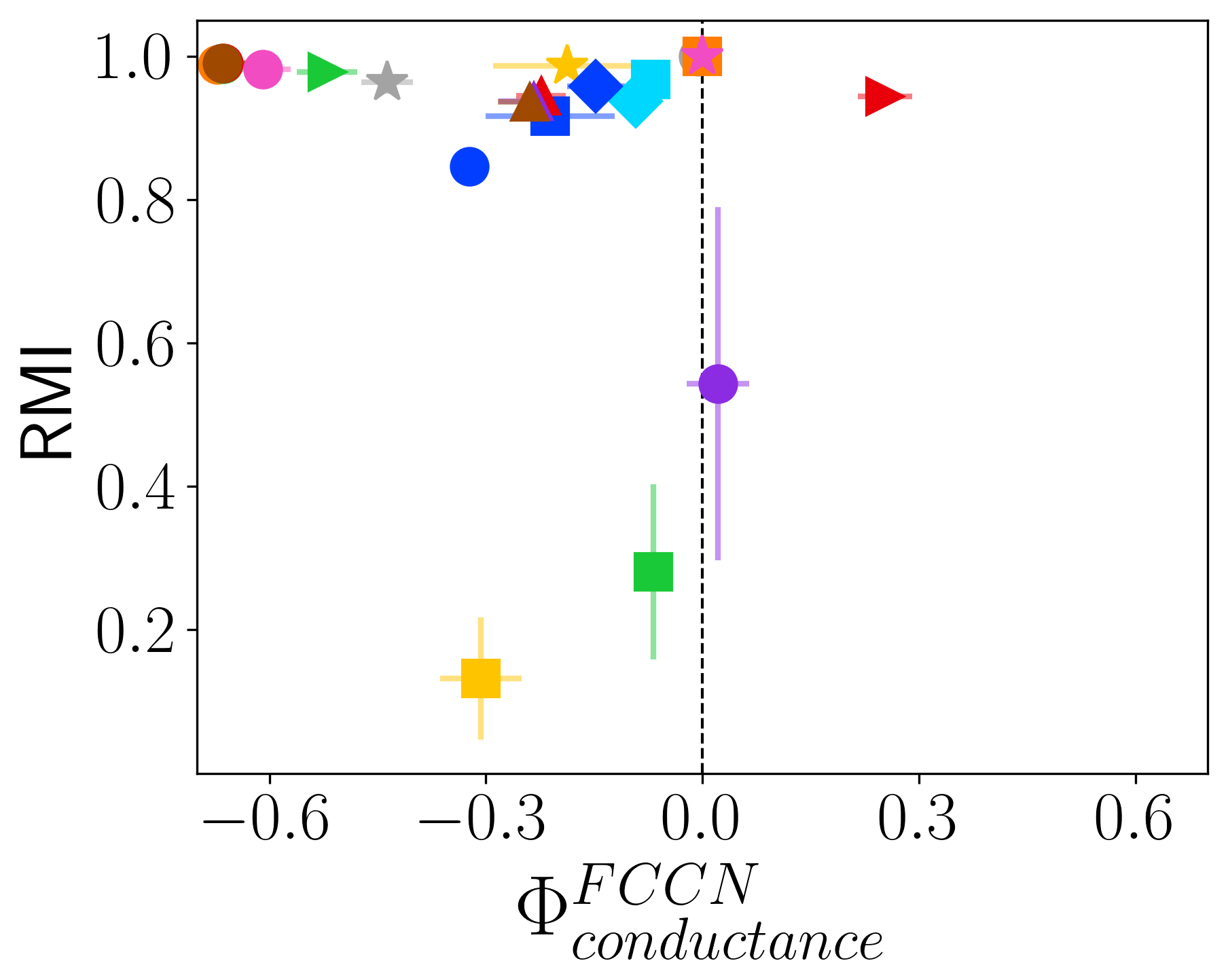}
\end{minipage}
\\
\begin{subfigure}[c]{0.05\textwidth}
\caption*{\rotatebox{90}{NF1}}
\end{subfigure}
\begin{minipage}[c]{0.94\textwidth}
\includegraphics[width=0.31\textwidth]{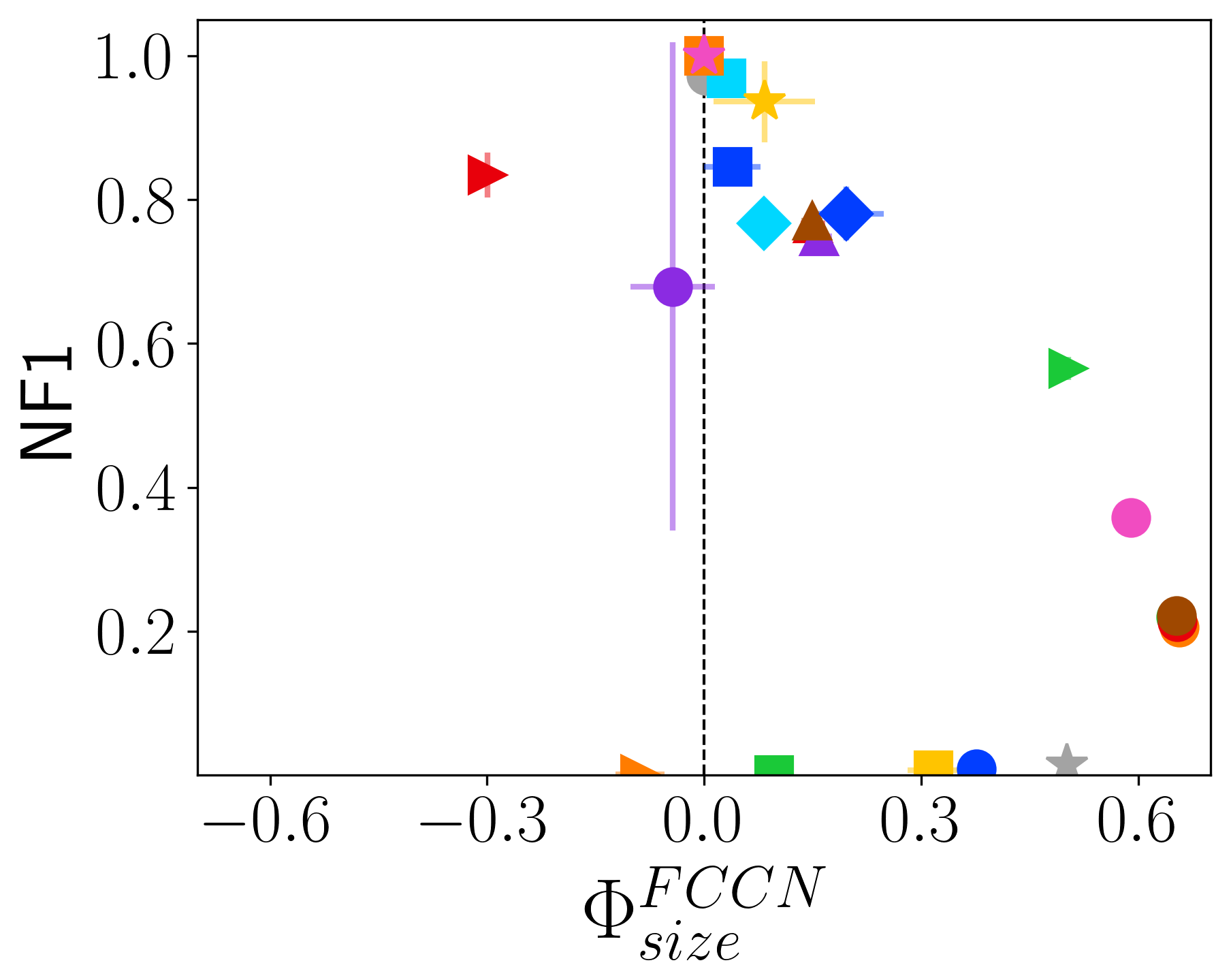}\quad
\includegraphics[width=0.31\textwidth]{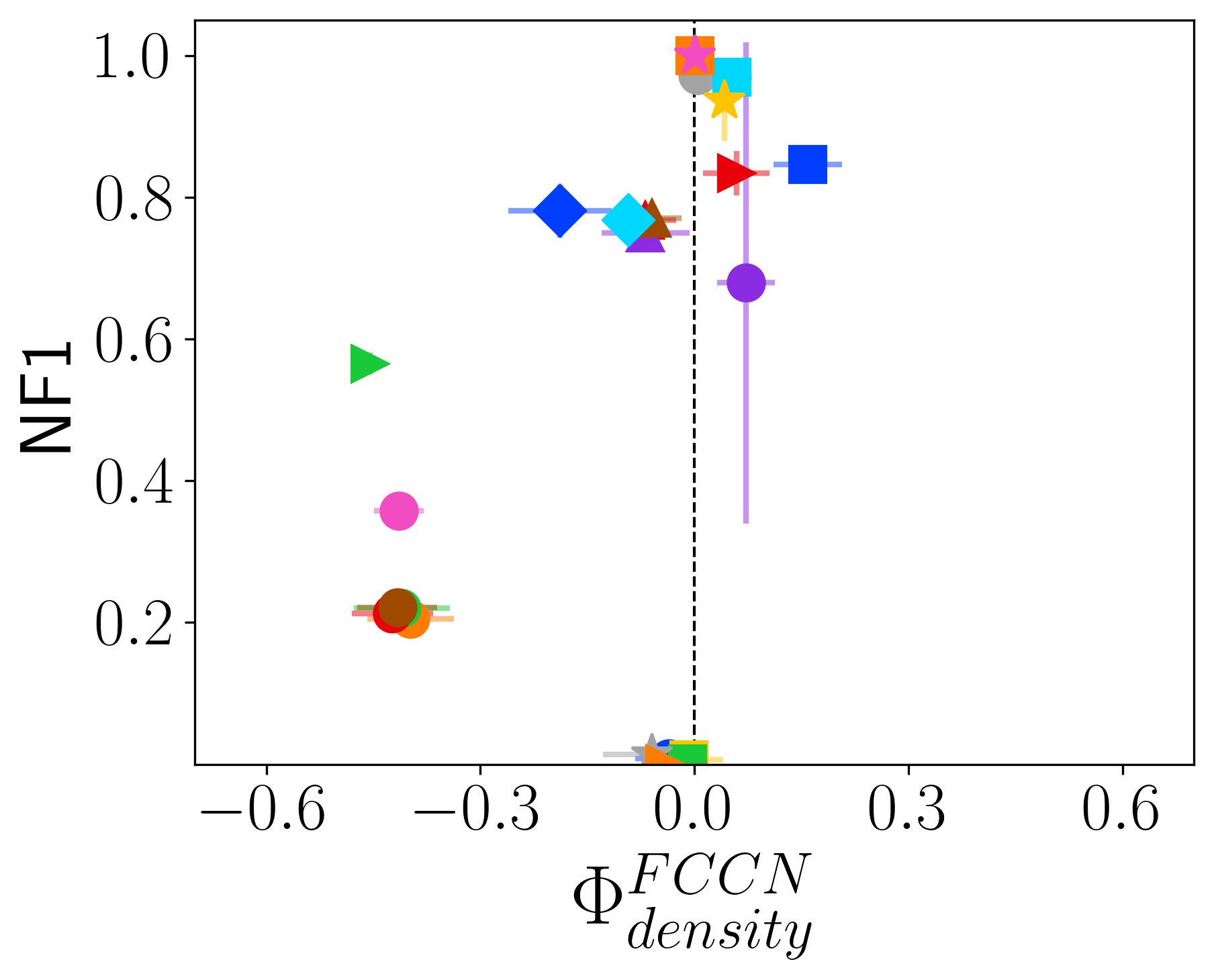}\quad
\includegraphics[width=0.31\textwidth]{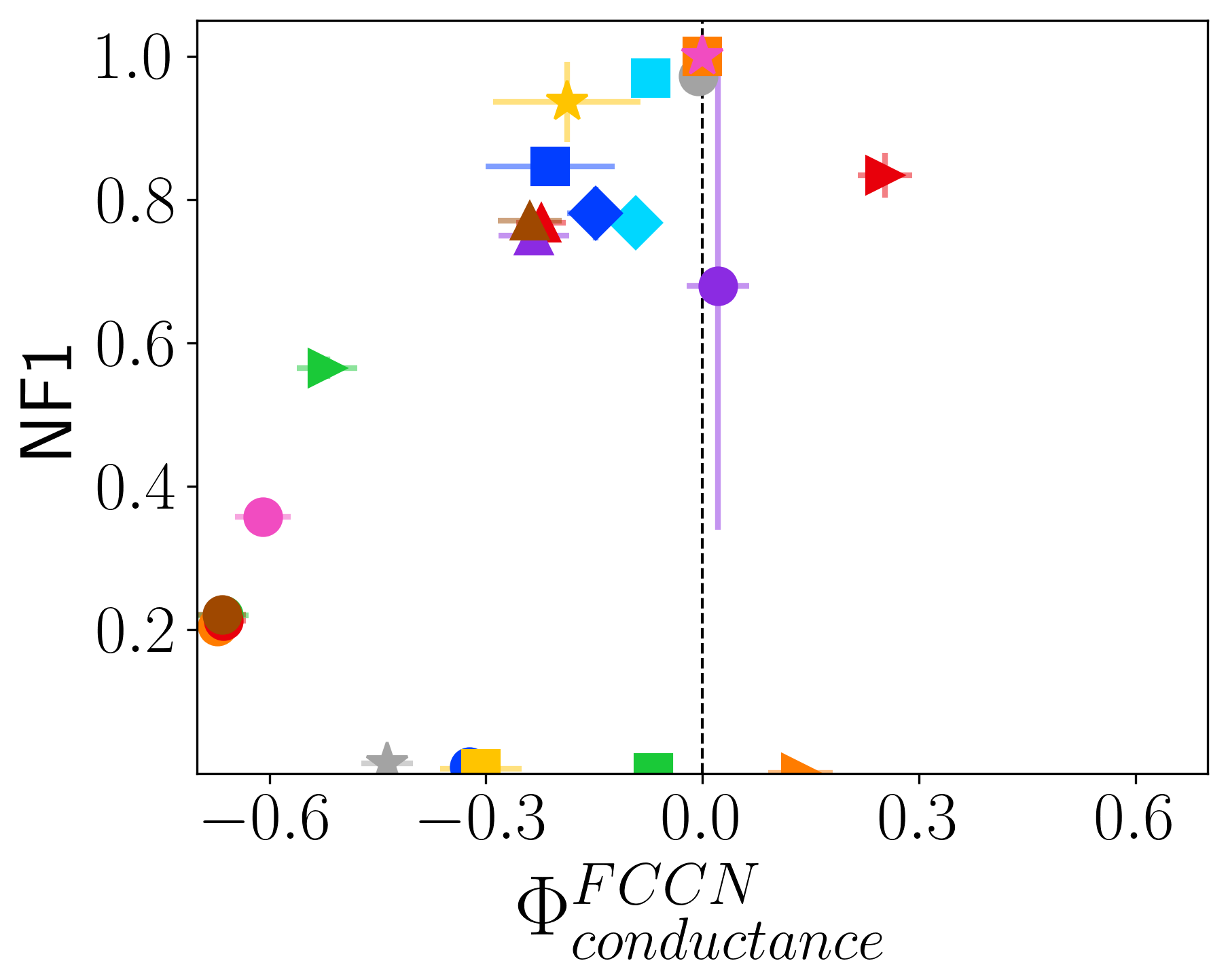}
\end{minipage}
\caption{ARI, RMI, and NF1 vs. $\Phi$ on LFR networks having $\mu=0.2$.}\label{phi_vs_other_metrics} 
\end{figure}

We also compare fairness with three additional quality evaluation metrics: (i) Adjusted Rand Index (ARI), (ii) Reduced Mutual Information (RMI), and (iii) Normalized F1 score (NF1) \cite{Chakraborty2016}. Fig. \ref{phi_vs_other_metrics} shows the results. The aim of this analysis was to identify CD methods which are both fair and perform well across various metrics, so that they can be recommended for broader use. Additionally, understanding the working dynamics of these methods will help in designing CD methods with high fairness-performance trade-off. From this analysis, the standout CD methods include: RSC-V, RSC-K, Walktrap, Infomap, and Significance. 

\subsection{Fairness Analysis on Real-world Networks}

\begin{figure}[t]
\centering
\begin{subfigure}[b]{0.93\textwidth}            
    \includegraphics[width=\textwidth]{legend_ncol6.png}
\end{subfigure}\\
\begin{subfigure}[c]{0.05\textwidth}
\caption*{\rotatebox{90}{Polbooks}}
\end{subfigure}%
\begin{minipage}[c]{0.95\textwidth}
\includegraphics[width=0.3\textwidth]{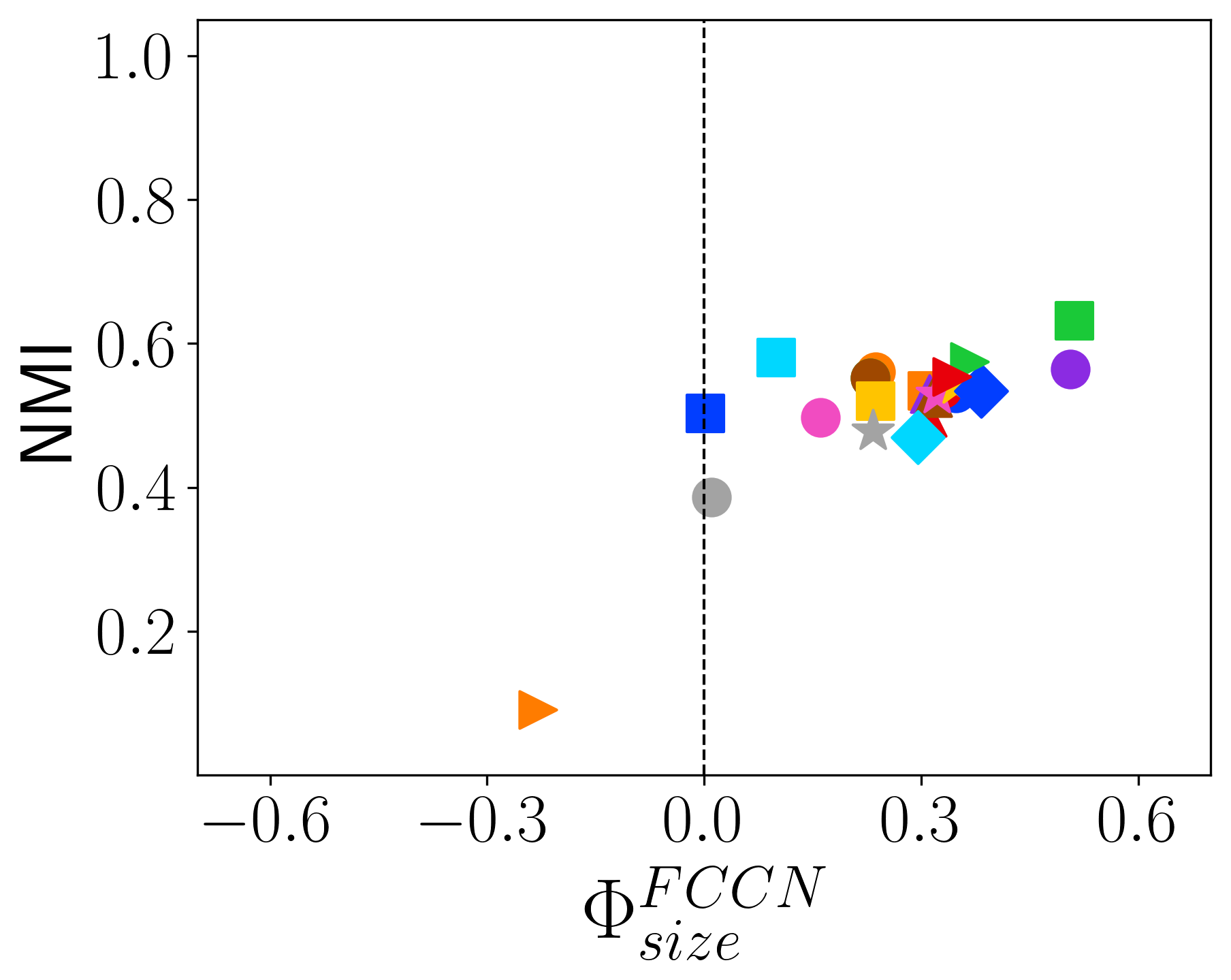}\quad
\includegraphics[width=0.3\textwidth]{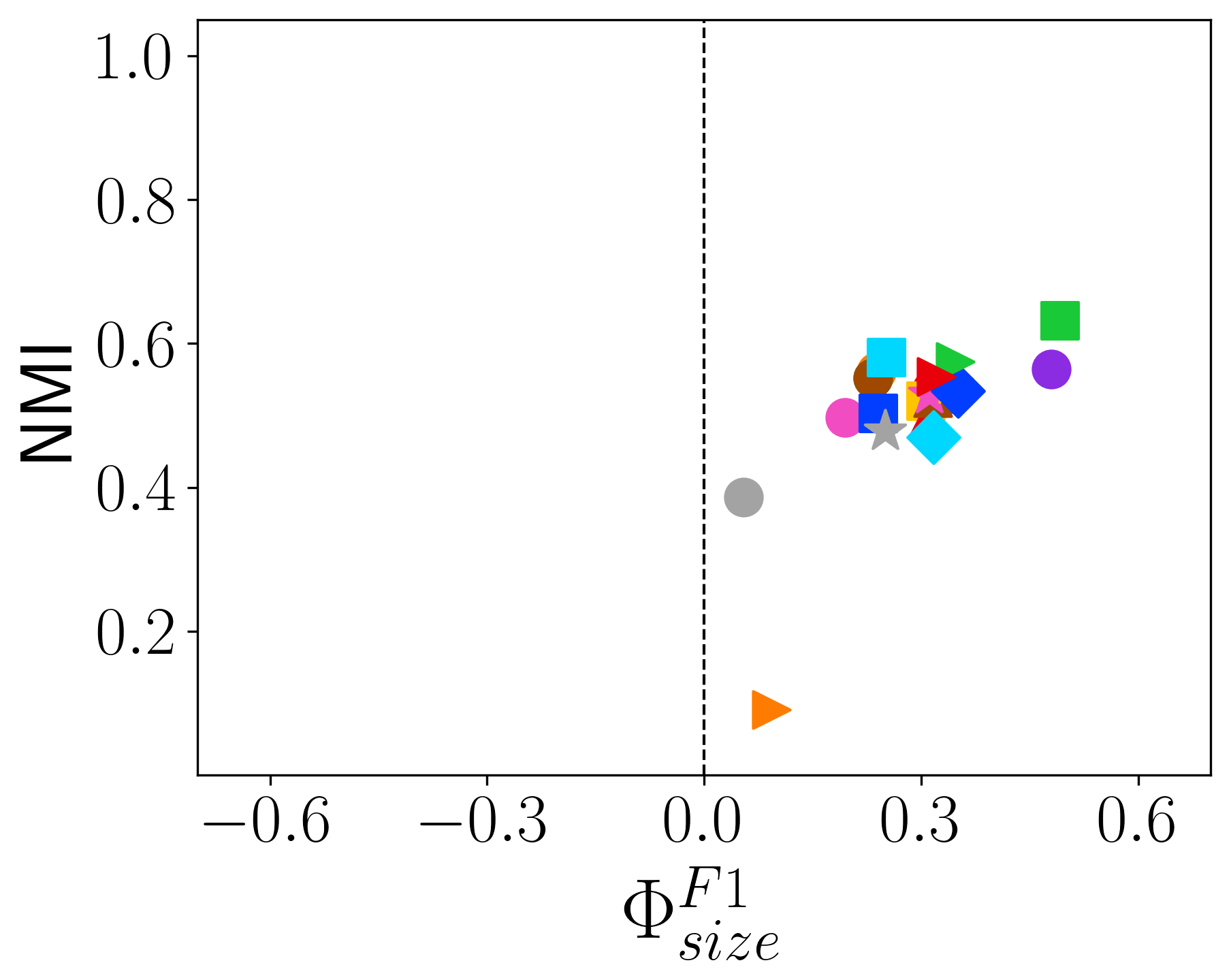}\quad
\includegraphics[width=0.3\textwidth]{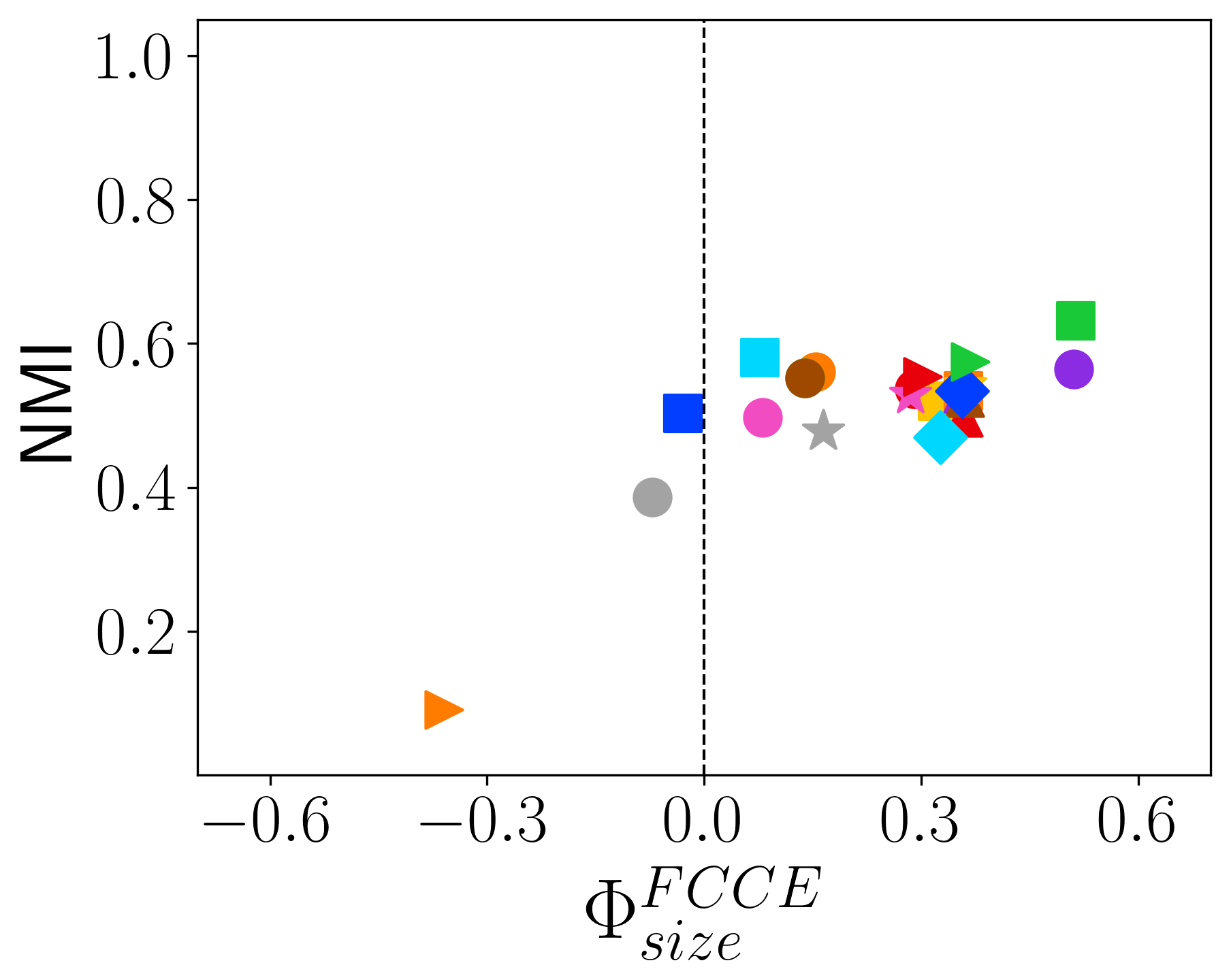}
\end{minipage}
\\
\begin{subfigure}[c]{0.05\textwidth}
\caption*{\rotatebox{90}{Football}}
\end{subfigure}%
\begin{minipage}[c]{0.95\textwidth}
\includegraphics[width=0.3\textwidth]{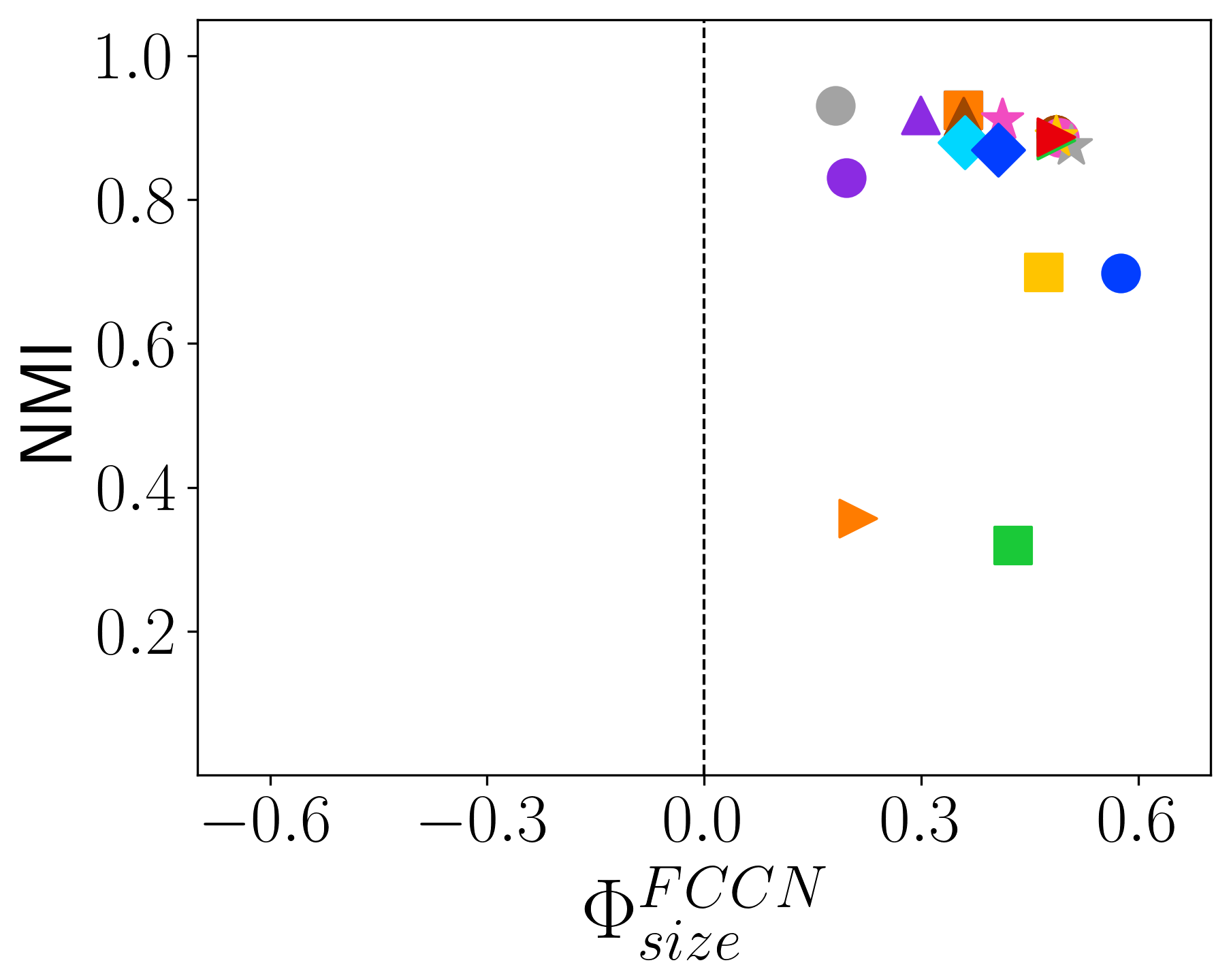}\quad
\includegraphics[width=0.3\textwidth]{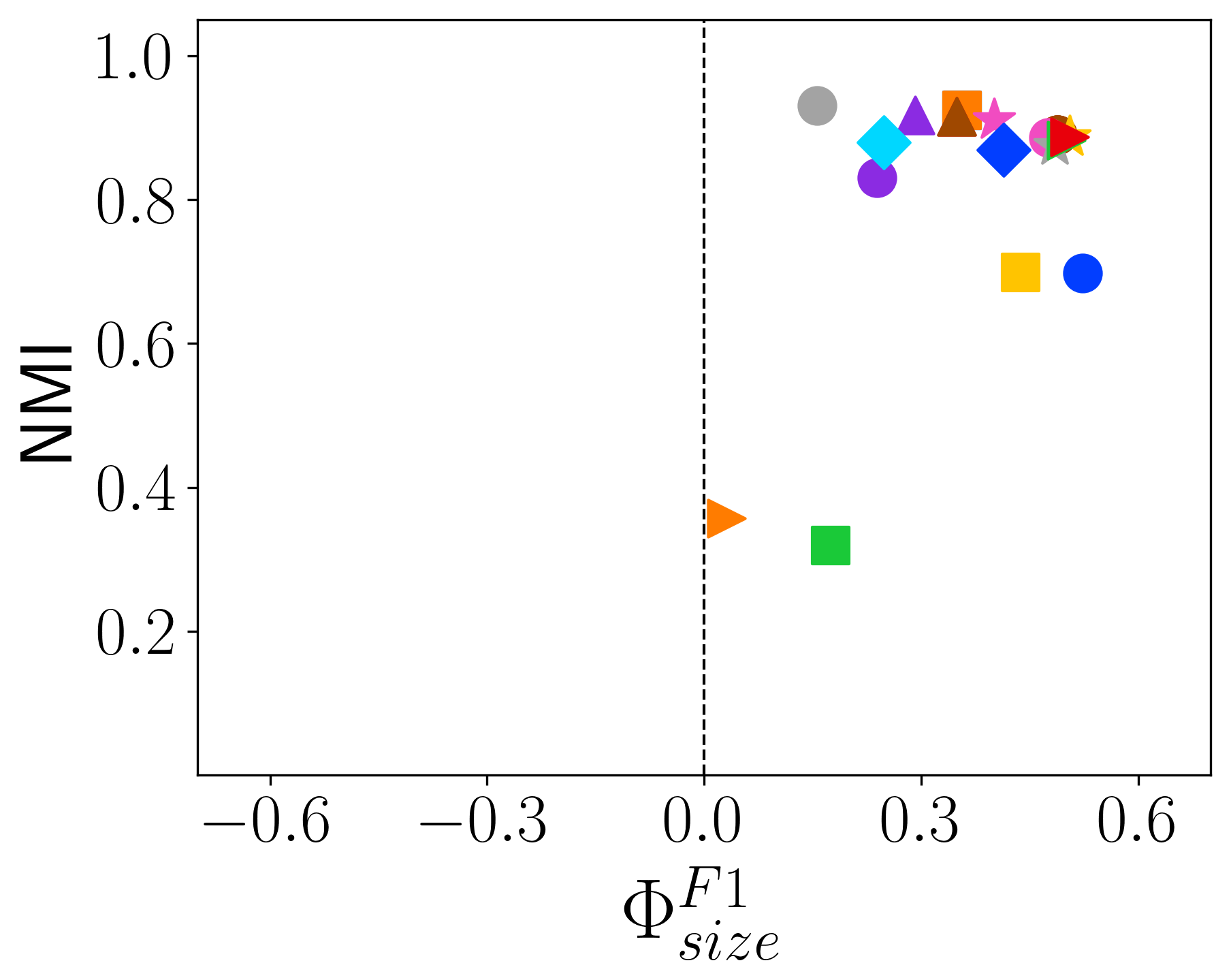}\quad
\includegraphics[width=0.3\textwidth]{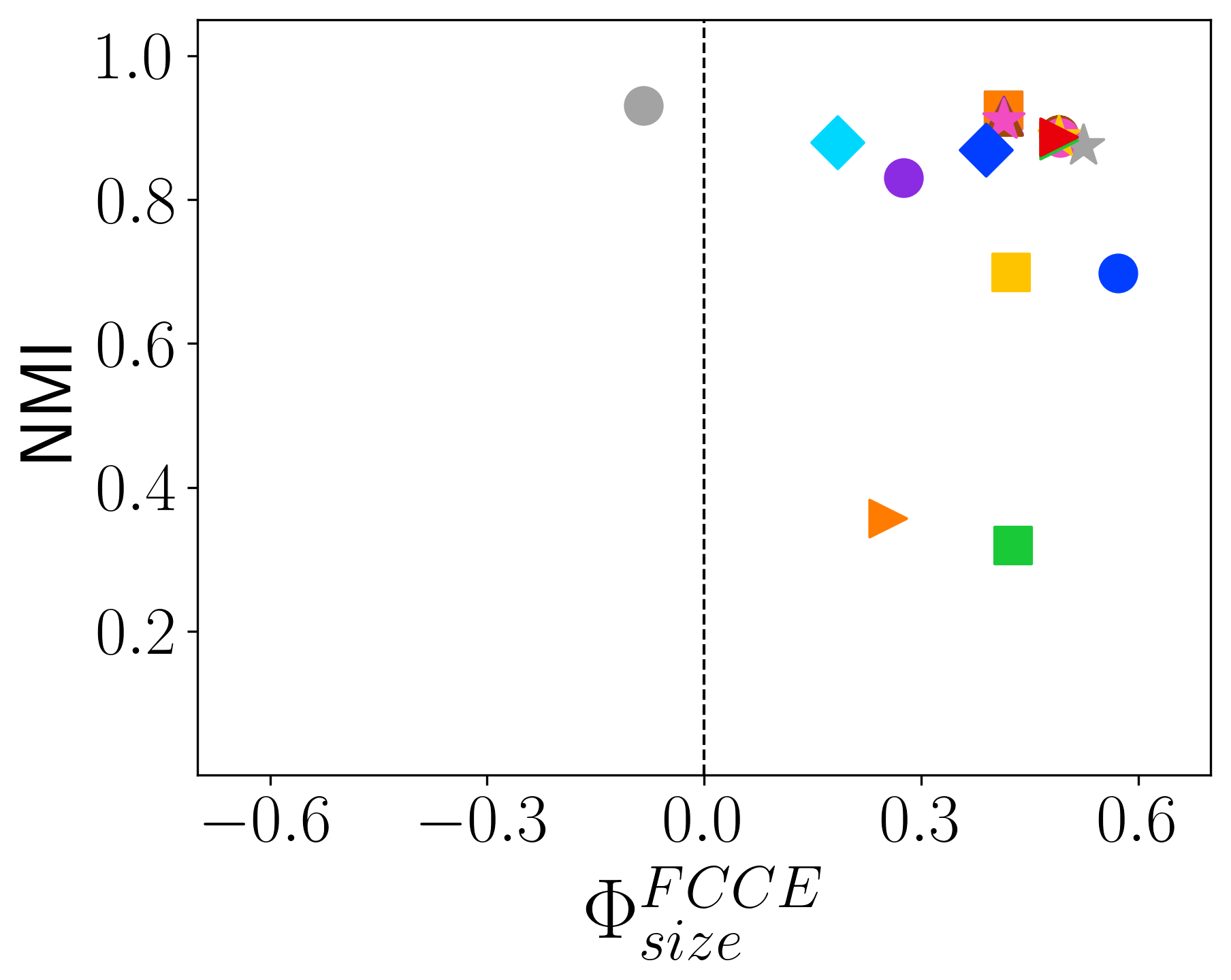}
\end{minipage}
\\
\begin{subfigure}[c]{0.05\textwidth}
\caption*{\rotatebox{90}{Eu-core}}
\end{subfigure}%
\begin{minipage}[c]{0.95\textwidth}
\includegraphics[width=0.3\textwidth]{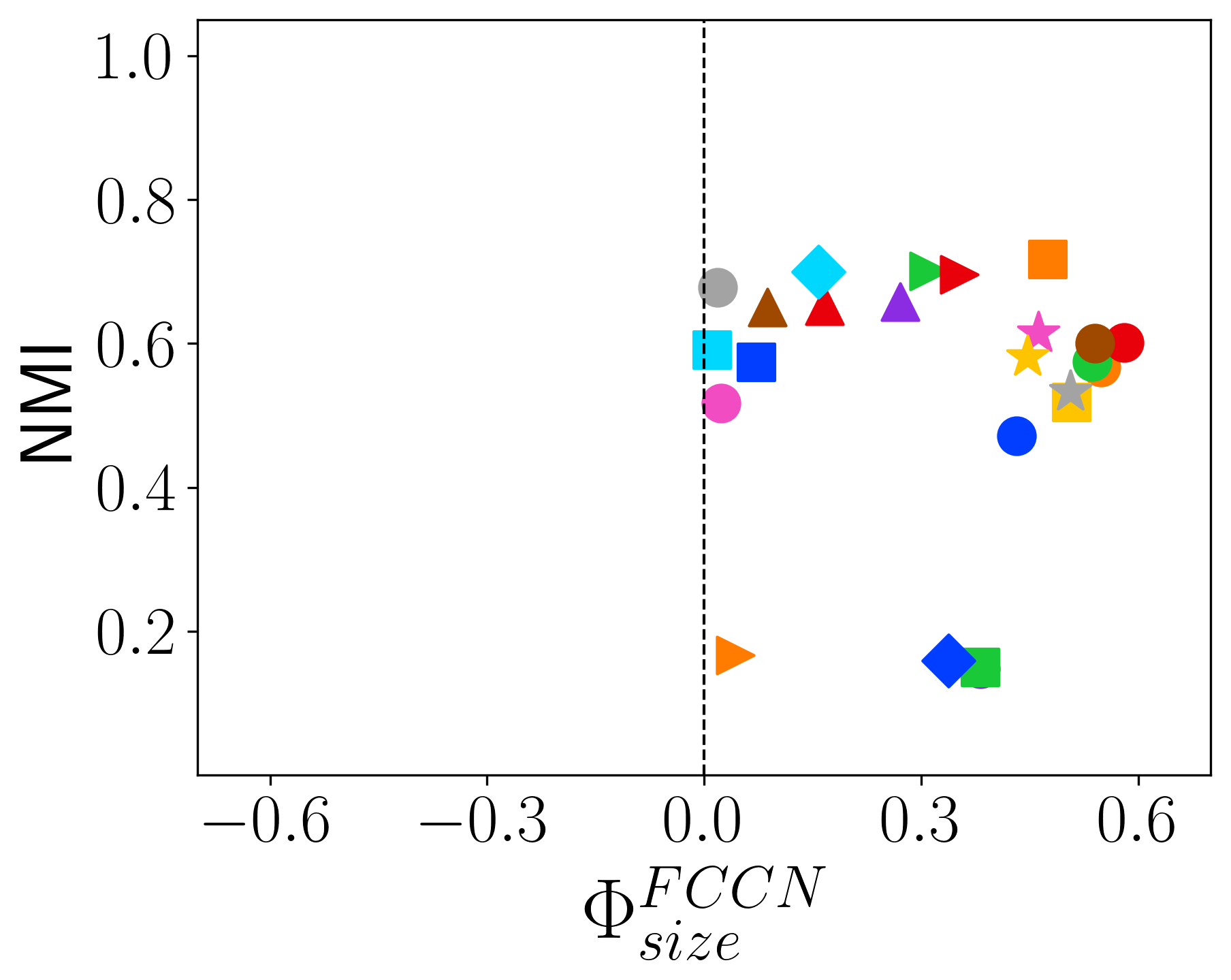}\quad
\includegraphics[width=0.3\textwidth]{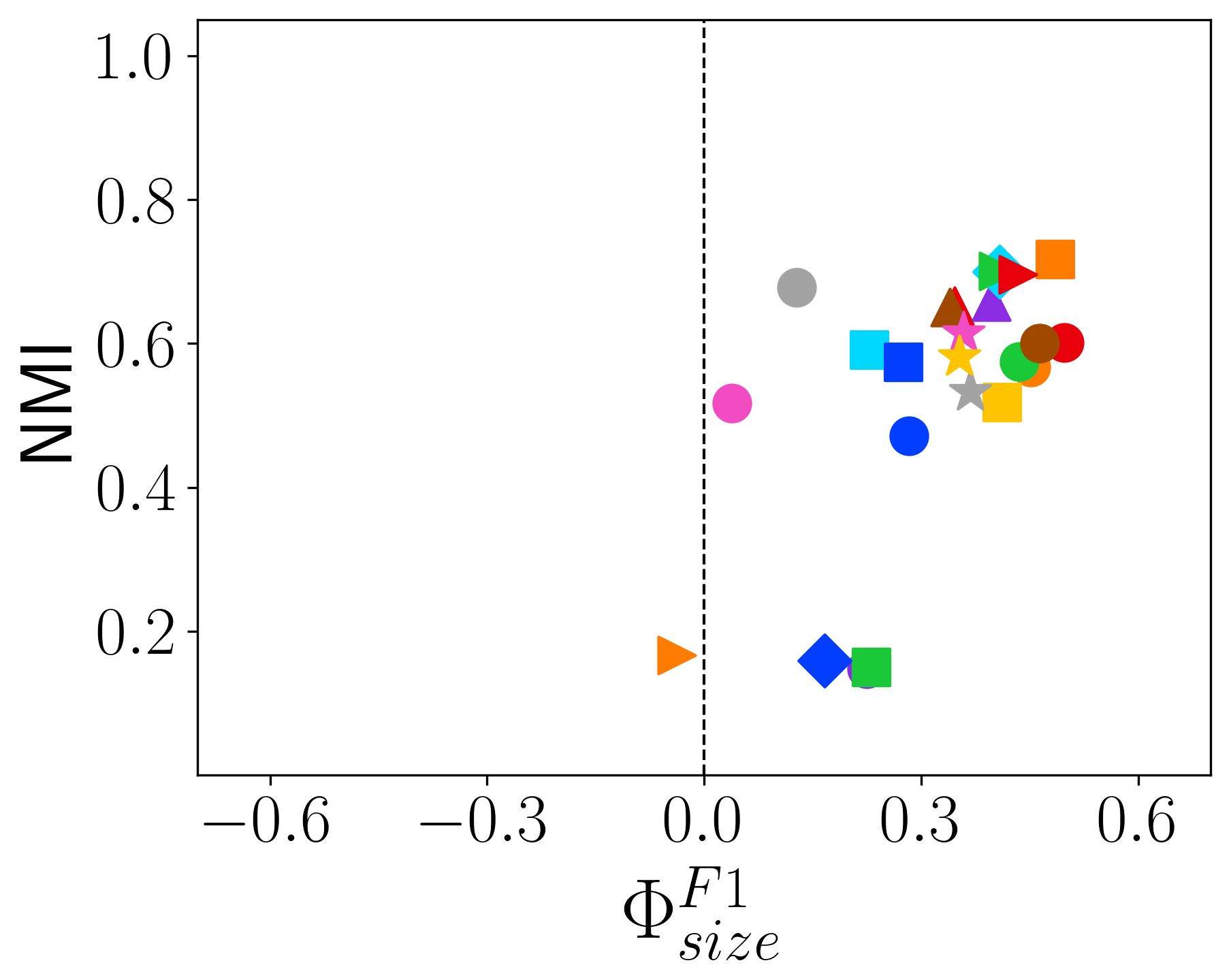}\quad
\includegraphics[width=0.3\textwidth]{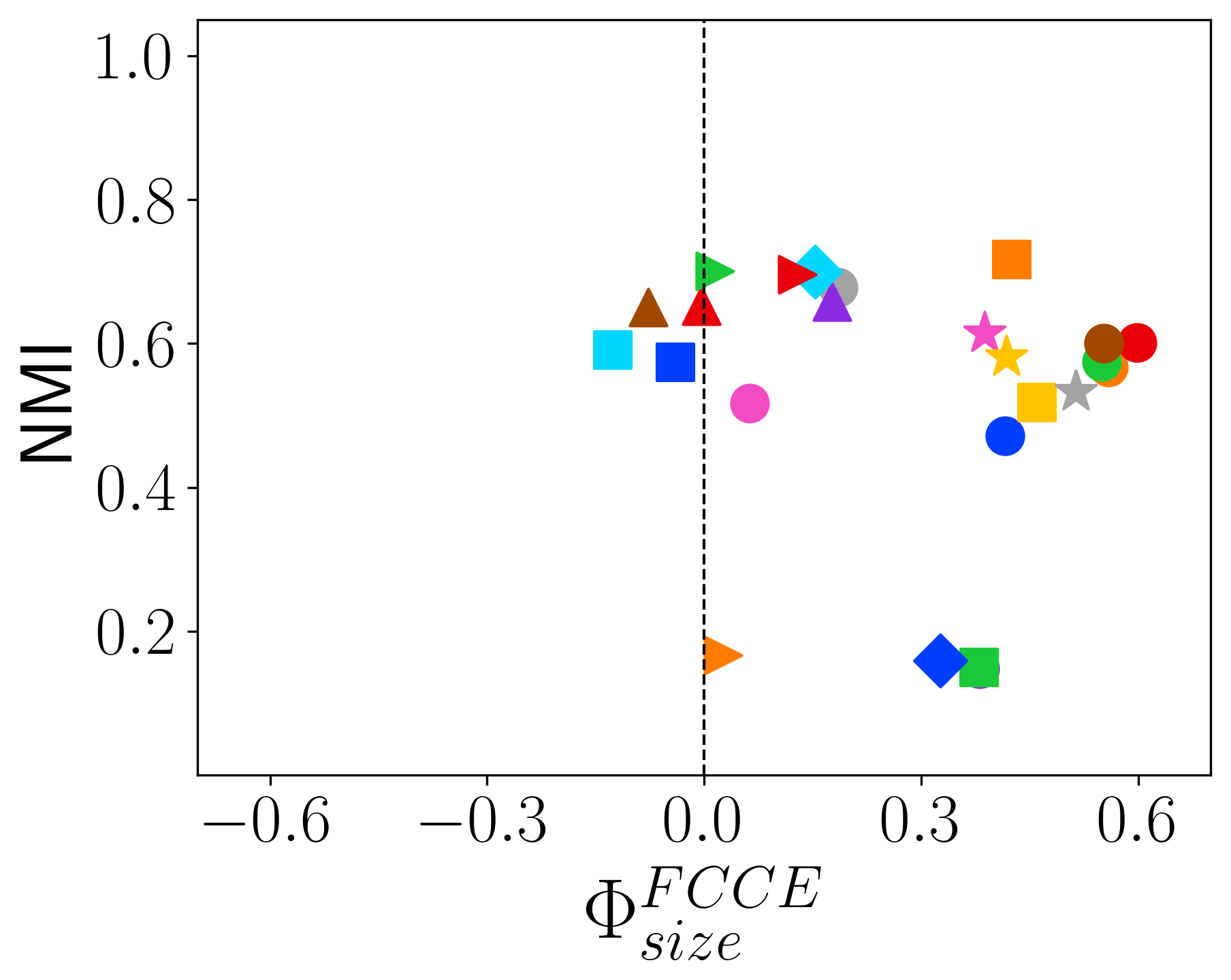}
\end{minipage}
\caption{NMI vs. fairness of CD methods on real-world networks.}\label{real_world_phi}
\end{figure}

Fig.~\ref{real_world_phi} presents NMI vs. $\Phi^{F*}_{size}$ for real-world networks. All methods, except EM, tend to favor larger groups, with consistent results across various fairness metrics. It is important to note that no single method is universally fair across all datasets, though Significance consistently performs well.

\section{Conclusion}\label{conclusion}

In real-world networks, communities vary in size, density, and connectivity due to how people organize and communicate. In this work, we first introduce community-wise fairness metrics (FCCN, F1, and FCCE) and use them to propose group fairness metrics ($\Phi$) to evaluate community detection methods with respect to communities' properties. We compare 24 community detection methods from six different classes on LFR benchmark and real-world networks. Our analysis reveals that no single class of methods consistently outperforms others; instead, performance varies for methods. However, some general trends emerge: modularity methods are highly biased towards large and sparse (except Paris and Significance), and low conductance groups. Representational and label propagation methods are biased toward large, low-density (for low $\mu$), and low-conductance groups. Dynamic and spectral methods have a huge variation in their fairness and quality evaluation. For $\mu=0.2$, the best methods which are fair in all aspects and have high quality scores (NMI, ARI, RMI, and NF1) include Significance, RSC-K, RSC-V, Infomap, and Walktrap. However, Walktrap is biased towards lower conductance. As $\mu$ increases, Fluid, SBM-Nested, SBM, and Label Propagation also perform better in all aspects, though they do not agree on RMI. 

This study offers valuable insights that can guide the design of fair community detection methods. In the future, we aim to propose individual fairness metrics for community detection to assess whether similar nodes are treated fairly. 

\bibliographystyle{spmpsci} 
\bibliography{refs} 
\end{document}